\documentclass[journal]{IEEEtran}
\usepackage{cite}
\usepackage{amsmath,amssymb,amsfonts}
\usepackage{algorithmic}
\usepackage{graphicx}
\usepackage{textcomp}
\usepackage{xcolor}
\usepackage{multicol}
\usepackage{multirow}
\usepackage[overload]{empheq}
\newcommand{\for}{\text{for }}

\begin{document}
\title{Learning-Driven Lossy Image Compression; A Comprehensive Survey}
\author{Sonain~Jamil,~Md.~Jalil~Piran,~\IEEEmembership{}and~MuhibUr~Rahman~\IEEEmembership{}
\thanks{S. Jamil is with the Department
of Electronics Engineering, Sejong University, Seoul, South Korea e-mail: sonainjamil@sju.ac.kr.}
\thanks{M. J. Piran is with the Department
of Computer Engineering, Sejong University, Seoul, South Korea e-mail: piran@sejong.ac.kr.}
\thanks{M. Rahman is with the Department
of Electrical Engineering, Polytechnique Montreal, Montreal, QC H3T 1J4, Canada e-mail: muhibur.rahman@polymtl.ca.}
\thanks{Manuscript received Month xx, xxxx; revised Month xx, xxxx.}}
\markboth{...}%
{S. Jamil \MakeLowercase{\textit{et al.}}: Learning-Driven Lossy Image Compression- A Survey}

\maketitle

\begin{abstract}
In the realm of image processing and computer vision (CV), machine learning (ML) architectures are widely applied. Convolutional neural networks (CNNs) solve a wide range of image processing issues and can solve image compression problem. Compression of images is necessary due to bandwidth and memory constraints. Helpful, redundant, and irrelevant information are three different forms of information found in images. This paper aims to survey recent techniques utilizing mostly lossy image compression using ML architectures including different auto-encoders (AEs) such as convolutional auto-encoders (CAEs), variational auto-encoders (VAEs), and AEs with hyper-prior models, recurrent neural networks (RNNs), CNNs, generative adversarial networks (GANs), principal component analysis (PCA) and fuzzy means clustering. We divide all of the algorithms into several groups based on architecture. We cover still image compression in this survey. Various discoveries for the researchers are emphasized and possible future directions for researchers. The open research problems such as out of memory (OOM), striped region distortion (SRD), aliasing, and compatibility of the frameworks with central processing unit (CPU) and graphics processing unit (GPU) simultaneously are explained. The majority of the publications in the compression domain surveyed are from the previous five years and use a variety of approaches. 
\end{abstract}
\begin{IEEEkeywords}
learned image compression, deep learning, JPEG, end-to-end image compression, machine learning.
\end{IEEEkeywords}


\section{Introduction}
\IEEEPARstart{I}{n} this modern era of big data, the size of the data is the biggest concern for scientists and researchers. Due to the limited bandwidth of the channel and limited memory space, there is the requirement of data compression for the successful transmission and storage of data without losing significant information \cite{b1}. Data compression can be performed in several ways: audio compression, image compression, video compression, and document compression \cite{bb2}. There are three types of information in an image: useful, redundant, and irrelevant. Irrelevant information can be ignored for the compression of images. Redundant information is crucial for highlighting details in images, whereas useful information is neither redundant nor irrelevant. Without accurate information, we cannot reconstruct or decompress images properly \cite{mm}. In image compression, there are two major categories. The first is lossless image compression, where no information is lost, and the second is lossy image compression. Lossless image compression techniques are very efficient for small-size data. 

Lossless techniques such as Huffman coding, run-length encoding (RLE), arithmetic coding, Lempel-Ziv-Welch (LZW) Coding, and JPEG-LS are efficient for the small data \cite{MARahman1}-\cite{MARahman2}. The major drawback of the lossless compression techniques is less compression efficiency than lossy compression techniques. That is why many researchers are working on image compression using ML. 

There are many surveys focused on image compression. Many surveys address the pros and cons of conventional image compression algorithms based on discrete cosine transform (DCT), and discrete wavelet transforms (DWT). Shukla \textit{et al.} in \cite{JShukla}, surveyed the prediction and transform-based image compression algorithms. This survey highlighted the importance and application of the prediction and transform-based conventional algorithms for image compression. These prediction-based algorithms were based on edge detection, gradient, and block-based prediction. However, transform-based algorithms were based on wavelets. The survey presented the comparative analysis of the entropy coding. However, this lacked the discussion about end-to-end based image compression architectures using ML. In \cite{Mehwish}, the authors studied the DCT and DWT-based algorithms. The survey did not address the recently learned image compression techniques. Likewise, in \cite{b2} the authors surveyed several lossy and lossless algorithms for image compression. The article highlighted the advantages and drawbacks of the predictive, entropy coding, and discrete Fourier transform-based image compression frameworks. The survey did not address the ML-based image compression architectures. Several surveys highlight the importance of the conventional DCT, DWT, and entropy-based approaches for image compression. There is a requirement to survey ML-based still image compression techniques.

A plethora of techniques have been proposed by researchers in recent years to address the problem of still image compression using ML techniques \cite{bb3}. So, there is a need to survey ML-based image compression techniques and highlight the open research problems. In this survey, we focus on recent ML-based image compression techniques. The following open research questions need to be addressed.
\begin{itemize}
	\item Which end-to-end still image compression using ML gives better compression efficiency?
	\item Which learned image compression frameworks give a better visual representation of the reconstructed image?
	\item Which compression technique saves GPU memory?
	\item Which compression technique has a fast response time?
\end{itemize}

This article presents a detailed survey of using ML algorithms for still image compression considering these questions. 
\begin{figure*}
    \centering
    \includegraphics[width=0.65\textwidth]{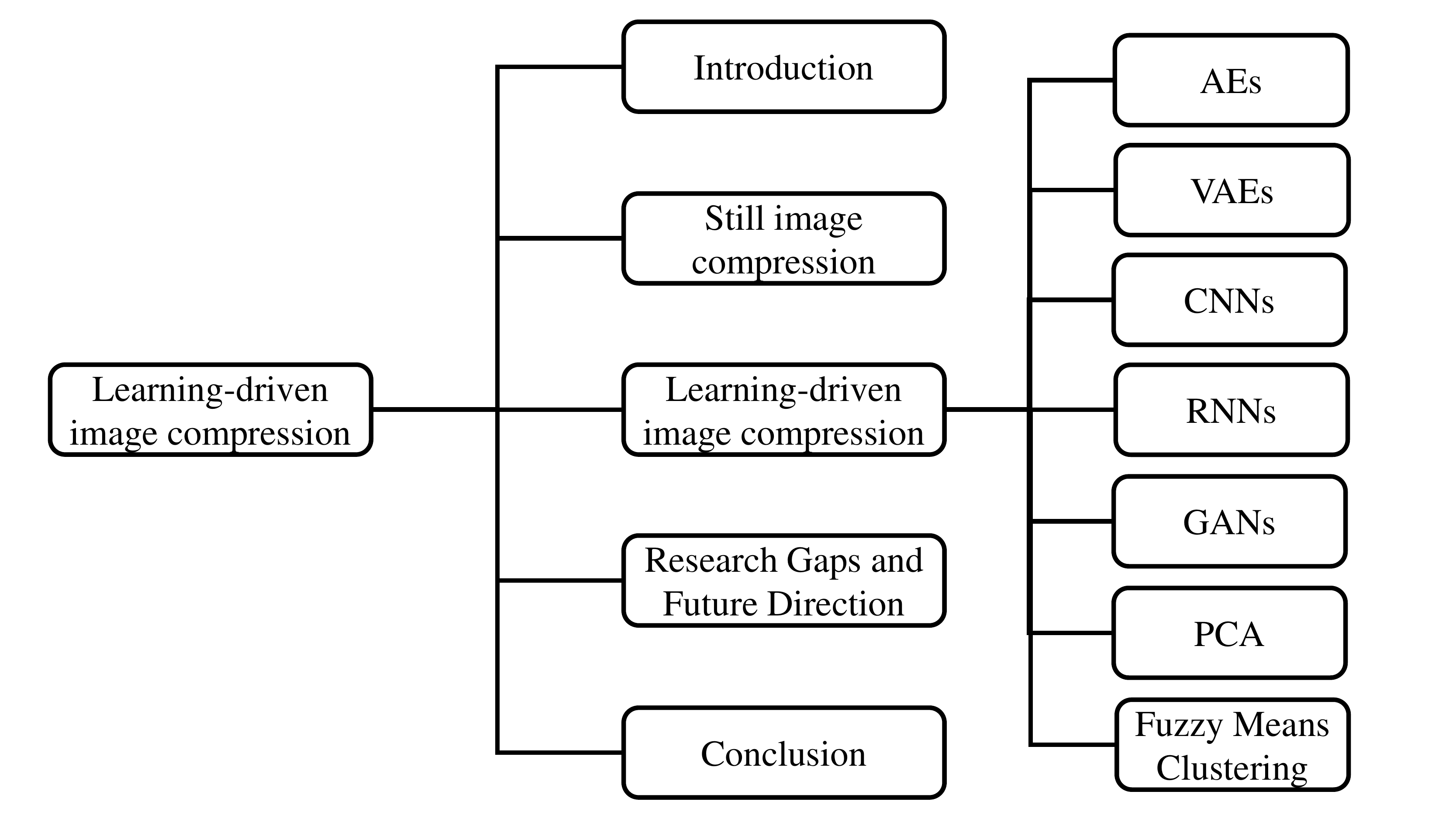}
    \caption{Organization of the paper.}
    \label{Taxonomy}
\end{figure*}

As presented in Fig. \ref{Taxonomy}, the rest of this paper is organized as follows. Section II presents the related work. In Section III, still image compression with conventional compression techniques is presented. The use of ML techniques for image compression is explained in Section IV. Section V gives an insight into the future research directions. Finally, Section VI draws the conclusions. 
\begin{table}
    \centering
    \caption{List of acronyms.}
    \begin{tabular}{p{40pt} p{190pt}}
    \hline
    \hline
    \textbf{Acronym}& \textbf{Meaning}\\
    \hline
    \hline
    AEs & Autoencoders\\
    BGAN+ & Binary Generative Adversarial Network\\
    BPG & Better Portable Graphics\\
    bpp & Bits per pixel\\
    CAEs & Convolutional Autoencoders\\
    CLIC & Challenge on Learned Image Compression\\
    CNNs & Convolutional Neural Networks\\
    C & Computer Vision\\
    CPU & Central Processing Unit\\
    DCT & Discrete Cosine Transform\\
    DNN & Deep Neural Network\\
    DWT & Discrete Wavelet Transform\\
    GANs & Generative Adversarial Networks\\
    GDN & Generalized Divisive Normalization\\
    GMM & Gaussian Mixture Module \\
    GPU & Graphics Processing Unit\\
    HDR & High Resolution\\
    HEVC & High Efficiency Video Coding\\
    IGDN & Inverse Generalized Divisive Normalization\\
    JPEG & Joint Photographic Expert Group\\
    JPEG XR & JPEG Extended Range\\
    KL & Kullback-Leiber\\
    LSTM & Long Short-Term Memory\\
    LZW & Lempel-Ziv-Welch\\
    ML & Machine Learning\\
    MNIST & Modified National Institute of Standards and Technology\\
    MSE & Mean Squared Error\\
    MS-SSIM & Multi-scale Structural Similarity Index Measure\\
    NL & Non-Local\\
    NLAM & Non-local Attention Module\\
    NLD & Non-linear Decoder\\
    NLE & Non-linear Encoder\\
    NN & Neural Network\\
    OOM & Out Of Memory\\
    PCA & Principal Component Analysis\\
    PNG & Portable Network Graphics\\
    PSNR & Peak Signal-to-Noise Ratio\\
    RD & Rate Distortion\\
    RLE & Run Length Encoding\\
    RNNs & Recurrent Neural Networks\\
    SCT & Stop Code Tolerant\\
    SRD & Striped Region Distortion\\
    SSIM & Structural Similarity Index Measure\\
    SWTA-AE & Stochastic Winner-Take-All Autoencoder\\
    VAEs & Variational Autoencoders\\
    VVC & Versatile Video Coding\\
    \hline
    \hline
    \end{tabular}
    \label{acronym}
\end{table}

\section{Related Work}
Still image compression has been a concern for researchers for many years. Initially, image compression was performed using conventional frameworks, and with the introduction of ML, algorithms used ML models for image compression. Several surveys focused on image compression algorithms. In \cite{Jiang1999}, the authors presented the comprehensive study of the image compression models based on the neural networks. This study lacked the discussion of other ML-based end-to-end image compression methods. The survey included old algorithms. Shum \textit{et al.} in \cite{Shum2003}, surveyed the DCT and DWT-based image compression and representation algorithms. The authors in \cite{JShukla} and \cite{Vrindavanam2012} conducted a study on conventional DCT and DWT-based frameworks. 

Similarly, the authors in \cite{Vijayvargiya2013} surveyed the entropy-based image compression algorithms. \cite{Mehwish} and \cite{Raid2014} surveyed the DCT and DWT-based image compression techniques in 2014. Likewise in 2017, Setyaningsih \textit{et al.} \cite{Setyaningsih2017}, surveyed the hybrid compression techniques. In \cite{b2}, the authors surveyed the predictive, entropy encoding, as well as discrete Fourier transform-based image compression architectures. In \cite{MARahman1} and \cite{MARahman2} the authors surveyed the conventional image compression algorithms. In \cite{MARahman1} the authors surveyed arithmetic entropy coding techniques while in \cite{MARahman2}, the authors presented the survey of image compression standards such as joint photographic experts group (JPEG), JPEG-2000, portable network graphics (PNG), WebP. Table \ref{TableSummary} presents the comprehensive summary of the related surveys.
\begin{table*}
    \centering
     \caption{Summary of the image compression surveys.}
    \begin{tabular}{p{50pt} p{50pt} p{40pt} p{40pt} p{40pt} p{40pt} p{40pt} p{100pt}}
    \hline
    \hline
         \vspace{0.01cm} \textbf{Survey} \vspace{0.1cm}& 
         \vspace{0.01cm} \textbf{Year} \vspace{0.1cm}& 
          \multicolumn{5}{p{200pt}}{\vspace{0.01cm} \textbf{Scope of the Architecture Surveyed} \vspace{0.1cm}}&
         \vspace{0.01cm} \textbf{Contributions and limitations} \vspace{0.1cm}\\\cline{3-7}
         \vspace{0.01cm} \vspace{0.1cm}& 
         \vspace{0.01cm} \vspace{0.1cm}& 
         \vspace{0.01cm} \textbf{AE} \vspace{0.1cm}& 
         \vspace{0.01cm} \textbf{VAE} \vspace{0.1cm}& 
         \vspace{0.01cm} \textbf{CNN} \vspace{0.1cm}& 
         \vspace{0.01cm} \textbf{RNN} \vspace{0.1cm}& 
         \vspace{0.01cm} \textbf{GAN} \vspace{0.1cm}& 
         \vspace{0.01cm} \textbf{} \vspace{0.1cm}\\ 
         \hline
         \vspace{0.01cm} \cite{Jiang1999} \vspace{0.1cm}& 
         \vspace{0.01cm} 1999 \vspace{0.1cm}& 
         \vspace{0.01cm} $\times$ \vspace{0.1cm}& 
         \vspace{0.01cm} $\times$ \vspace{0.1cm}& 
         \vspace{0.01cm} $\times$ \vspace{0.1cm}& 
         \vspace{0.01cm} $\times$ \vspace{0.1cm}& 
         \vspace{0.01cm} $\times$ \vspace{0.1cm}&  
         \vspace{0.01cm} Survey of NN based frameworks only \vspace{0.1cm}\\
         \hline
         \vspace{0.01cm} \cite{Shum2003} \vspace{0.1cm}& 
		\vspace{0.01cm} 2003 \vspace{0.1cm}& 
		\vspace{0.01cm} $\times$ \vspace{0.1cm}& 
         \vspace{0.01cm} $\times$ \vspace{0.1cm}& 
         \vspace{0.01cm} $\times$ \vspace{0.1cm}& 
         \vspace{0.01cm} $\times$ \vspace{0.1cm}& 
         \vspace{0.01cm} $\times$ \vspace{0.1cm}& \vspace{0.01cm} Survey of DCT and DWT-based image representation and compression algorithms \vspace{0.1cm}\\
		\hline
         \vspace{0.01cm} \cite{JShukla} \vspace{0.1cm}& 
         \vspace{0.01cm} 2010 \vspace{0.1cm}& 
         \vspace{0.01cm} $\times$ \vspace{0.1cm}& 
         \vspace{0.01cm} $\times$ \vspace{0.1cm}& 
         \vspace{0.01cm} $\times$ \vspace{0.1cm}& 
         \vspace{0.01cm} $\times$ \vspace{0.1cm}& 
         \vspace{0.01cm} $\times$ \vspace{0.1cm}& \vspace{0.01cm} Survey of predictive and transform based image compression methods \vspace{0.1cm}\\
         \hline
         \vspace{0.01cm} \cite{Vrindavanam2012} \vspace{0.1cm}& 
         \vspace{0.01cm} 2012 \vspace{0.1cm}& 
         \vspace{0.01cm} $\times$ \vspace{0.1cm}& 
         \vspace{0.01cm} $\times$ \vspace{0.1cm}& 
         \vspace{0.01cm} $\times$ \vspace{0.1cm}& 
         \vspace{0.01cm} $\times$ \vspace{0.1cm}& 
         \vspace{0.01cm} $\times$ \vspace{0.1cm}& \vspace{0.01cm} Survey of entropy based encoding, JPEG, JPEG2000 methods \vspace{0.1cm}\\
         \hline
         \vspace{0.01cm} \cite{Vijayvargiya2013} \vspace{0.1cm}& 
         \vspace{0.01cm} 2013 \vspace{0.1cm}& 
         \vspace{0.01cm} $\times$ \vspace{0.1cm}& 
         \vspace{0.01cm} $\times$ \vspace{0.1cm}& 
         \vspace{0.01cm} $\times$ \vspace{0.1cm}& 
         \vspace{0.01cm} $\times$ \vspace{0.1cm}& 
         \vspace{0.01cm} $\times$ \vspace{0.1cm}& \vspace{0.01cm} Survey of entropy based encoding methods \vspace{0.1cm}\\
         \hline
         \vspace{0.01cm} \cite{Mehwish} \vspace{0.1cm}& 
         \vspace{0.01cm} 2014 \vspace{0.1cm}& 
         \vspace{0.01cm} $\times$ \vspace{0.1cm}& 
         \vspace{0.01cm} $\times$ \vspace{0.1cm}& 
         \vspace{0.01cm} $\times$ \vspace{0.1cm}& 
         \vspace{0.01cm} $\times$ \vspace{0.1cm}& 
         \vspace{0.01cm} $\times$ \vspace{0.1cm}& \vspace{0.01cm} Survey of DCT and DWT based architectures only \vspace{0.1cm}\\
         \hline
         \vspace{0.01cm} \cite{Raid2014} \vspace{0.1cm}& 
         \vspace{0.01cm} 2014 \vspace{0.1cm}& 
         \vspace{0.01cm} $\times$ \vspace{0.1cm}& 
         \vspace{0.01cm} $\times$ \vspace{0.1cm}& 
         \vspace{0.01cm} $\times$ \vspace{0.1cm}& 
         \vspace{0.01cm} $\times$ \vspace{0.1cm}& 
         \vspace{0.01cm} $\times$ \vspace{0.1cm}& \vspace{0.01cm} Survey of JPEG and DCT based architectures \vspace{0.1cm}\\
         \hline
         \vspace{0.01cm} \cite{Setyaningsih2017} \vspace{0.1cm}& 
         \vspace{0.01cm} 2017 \vspace{0.1cm}& 
         \vspace{0.01cm} $\times$ \vspace{0.1cm}& 
         \vspace{0.01cm} $\times$ \vspace{0.1cm}& 
         \vspace{0.01cm} $\times$ \vspace{0.1cm}& 
         \vspace{0.01cm} $\times$ \vspace{0.1cm}& 
         \vspace{0.01cm} $\times$ \vspace{0.1cm}& \vspace{0.01cm} Survey of hybrid DCT, DWT based compression techniques \vspace{0.1cm}\\
         \hline
         \vspace{0.01cm} \cite{b2} \vspace{0.1cm}& 
         \vspace{0.01cm} 2018 \vspace{0.1cm}& 
         \vspace{0.01cm} $\times$ \vspace{0.1cm}& 
         \vspace{0.01cm} $\times$ \vspace{0.1cm}& 
         \vspace{0.01cm} $\times$ \vspace{0.1cm}& 
         \vspace{0.01cm} $\times$ \vspace{0.1cm}& 
         \vspace{0.01cm} $\times$ \vspace{0.1cm}& \vspace{0.01cm} Survey of predictive, entropy coding and discrete Fourier transform based architectures \vspace{0.1cm}\\
         \hline
         \vspace{0.01cm} \cite{MARahman1} \vspace{0.1cm}& 
         \vspace{0.01cm} 2019 \vspace{0.1cm}& 
         \vspace{0.01cm} $\times$ \vspace{0.1cm}& 
         \vspace{0.01cm} $\times$ \vspace{0.1cm}& 
         \vspace{0.01cm} $\times$ \vspace{0.1cm}& 
         \vspace{0.01cm} $\times$ \vspace{0.1cm}& 
         \vspace{0.01cm} $\times$ \vspace{0.1cm}& \vspace{0.01cm} Survey of arithmetic encoding based architectures only \vspace{0.1cm}\\
         \hline
         \vspace{0.01cm} \cite{MARahman2} \vspace{0.1cm}& 
         \vspace{0.01cm} 2021 \vspace{0.1cm}& 
         \vspace{0.01cm} $\times$ \vspace{0.1cm}& 
         \vspace{0.01cm} $\times$ \vspace{0.1cm}& 
         \vspace{0.01cm} $\times$ \vspace{0.1cm}& 
         \vspace{0.01cm} $\times$ \vspace{0.1cm}& 
         \vspace{0.01cm} $\times$ \vspace{0.1cm}& \vspace{0.01cm} Survey of conventional JPEG, JPEG2000, WebP, PNG based architectures \vspace{0.1cm}\\
         \hline
         \vspace{0.01cm} This survey \vspace{0.1cm}& 
         \vspace{0.01cm} 2021 \vspace{0.1cm}& 
         \vspace{0.01cm} \checkmark \vspace{0.1cm}& 
         \vspace{0.01cm} \checkmark \vspace{0.1cm}& 
         \vspace{0.01cm} \checkmark \vspace{0.1cm}& 
         \vspace{0.01cm} \checkmark \vspace{0.1cm}& 
         \vspace{0.01cm} \checkmark \vspace{0.1cm}& \vspace{0.01cm} Survey of end-to-end ML based image compression frameworks, New outlook to the open research gaps \vspace{0.1cm}\\
         \hline
         \hline
     \end{tabular}
     \label{TableSummary}
\end{table*}

After analyzing Table \ref{TableSummary}, it is evident that all the existing surveys deal with the conventional compression techniques. However, there is a plethora of learning-driven lossy image compression frameworks and these architectures have limitations, pros and cons. There is a need for to survey learning-driven lossy image compression models.In this survey, we provide an in-depth detailed discussion regarding still lossy image compression. We present analysis, comparison, pros and cons of  the state-of-the-art techniques used for still lossy image compression. We outline several open research problems and trends in this research field for substantial future research.

The upcoming section explains the still image compression.
\section{Still Image Compression}
\begin{figure*}
    \centering
    \includegraphics[width=0.6\textwidth]{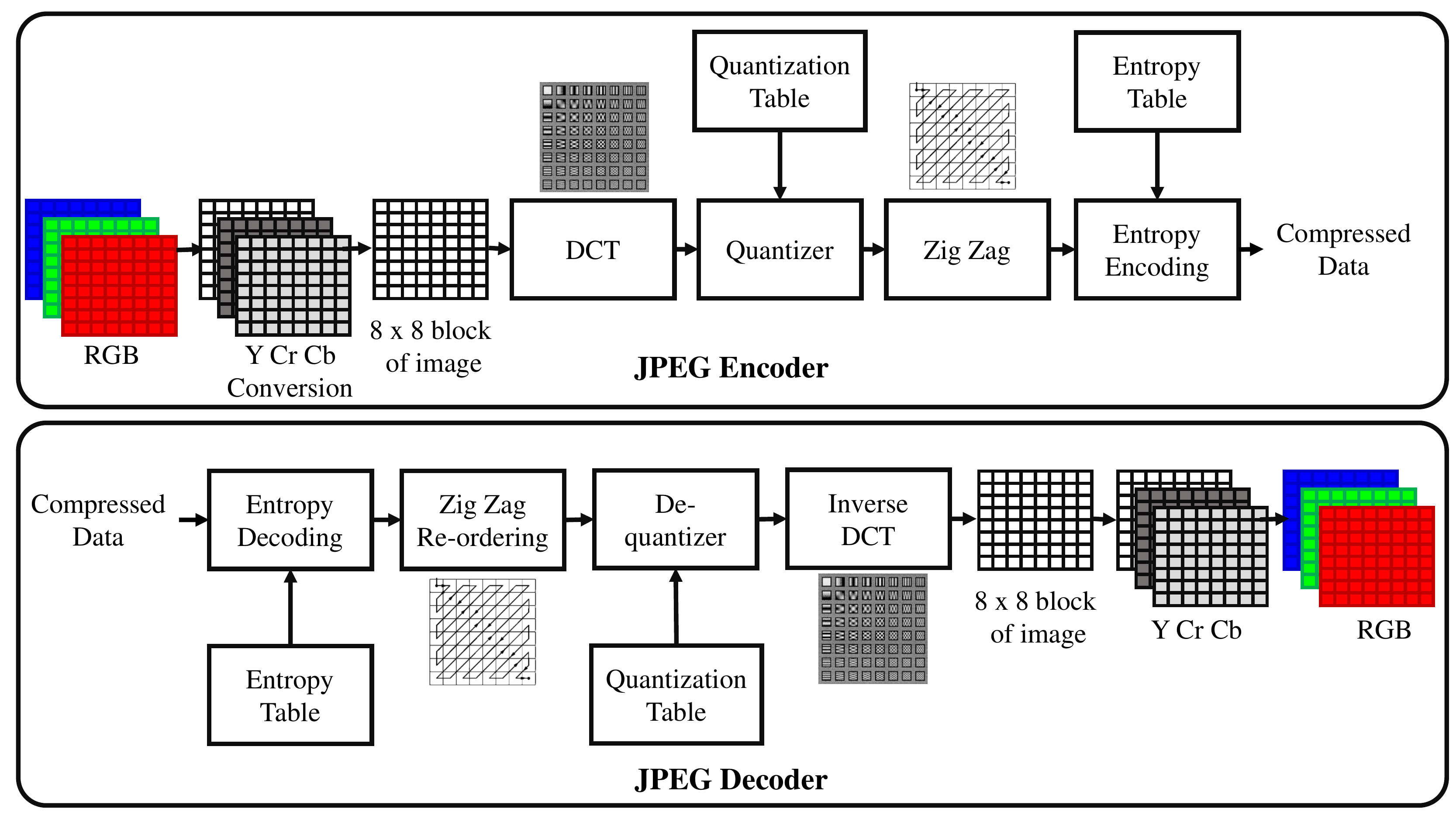}
    \caption{Block diagram of JPEG compression and decompression.}
    \label{JPEG}
\end{figure*}

The concept of image compression started in 1992 when Wallace proposed the JPEG standard for still image compression \cite{b3}. In the JPEG compression standard, DCT is used, which can be calculated as follows \cite{b4}.
\begin{equation}
    F(u,v)=\frac{1}{\sqrt{2N}}C(u)C(v)\sum_{x=0}^{N-1}\sum_{y=0}^{N-1}f(x,y) \dots\hfill \atop cos(\frac{(2x+1)u\pi}{N})cos(\frac{(2y+1)v\pi}{N}),
    \label{e1}
\end{equation}
where $x$ is the pixel row, for the integer $0\leqslant x \leqslant N$, $y$ is the pixel column, for the integer $0\leqslant y \leqslant N$, $F_{u,v}$ is the reconstructed approximate coefficient at coordinates $(u,v)$, $f_{x,y}$ is the reconstructed pixel value at coordinates $(x,y)$ and $C(u)$ is 
\begin{equation}
    C(u)=
    \begin{cases}
      \frac{1}{\sqrt{N}},& \for u=0 \\
      1,& \for otherwise.
    \end{cases}
\label{e2}
\end{equation}
The basic framework of JPEG compression and decompression is shown in Fig. \ref{JPEG}.

\begin{figure*}
    \centering
    \includegraphics[width=0.6\textwidth]{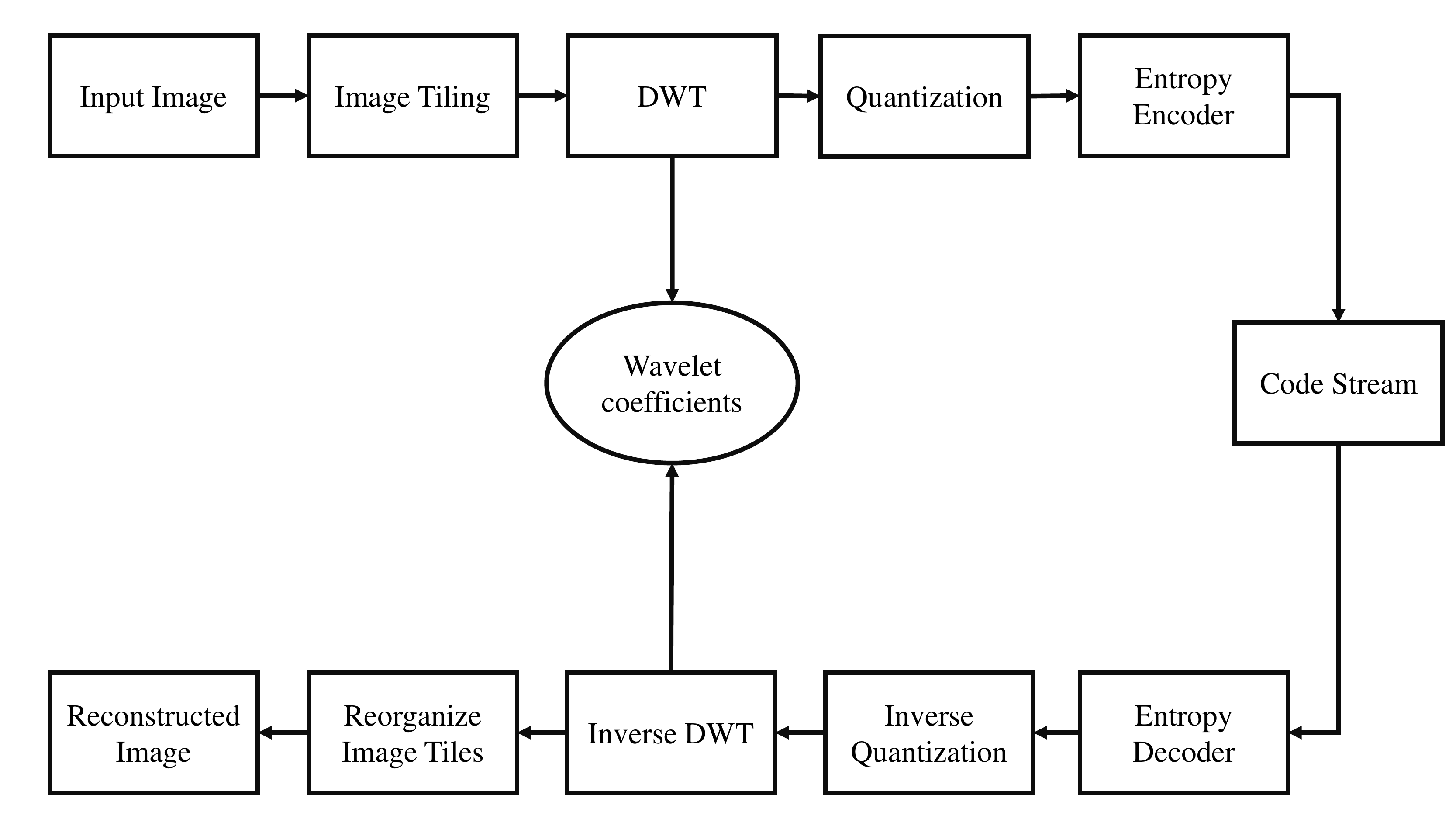}
    \caption{Block diagram of JPEG-2000 compression and decompression.}
    \label{JPEG2000}
\end{figure*}

JPEG was widely used for still image compression until an updated, and improved version of JPEG known as JPEG-2000 was proposed in 2001 \cite{b5}. JPEG-2000 uses DWT instead of DCT. The basic framework of JPEG-2000 is shown in Fig. \ref{JPEG2000}. The entropy encoding scheme used in JPEG-2000 is Huffman encoding \cite{b6}, RLE \cite{b7} or arithmetic encoding \cite{b8}.   

JPEG and JPEG-2000 standards have been widely used. Later on, researchers developed JPEG extended range (JPEG XR) \cite{b9}, and JPEG XT \cite{b10} to overcome the limitations of the JPEG standard. However, both these algorithms failed due to hardware incompatibility. Recently, WebP \cite{b11}, better portable graphics (BPG) \cite{b12} and JPEG XL \cite{b13} algorithm were proposed for the compression of high resolution (HDR) images. Recently, high-efficiency video coding (HEVC) displayed the most impressive performance. In \cite{HEVC1}, the authors explained the use of HEVC for high-depth medical image compression. They also proposed a novel algorithm to reduce the complexity of HEVC. In \cite{b14}, the authors presented a comparative analysis of all these algorithms. These conventional algorithms still have limitations. ML has addressed several issues like visual quality improvement for image compression. The next section presents learning-driven lossy image compression techniques. 
\section{Learning-Driven Lossy Image Compression}
ML has a significant influence on every field of research. There has been a revolution in image processing due to CNN's compiling property of features extractions. Inspired by this, many ML architectures have been proposed to compress images. We categorize architectures concerning ML models.
\subsection{Compression with AE}
The most widely used learning-driven lossy image compression model was proposed by Ball{\'e} in 2016 \cite{b15}. The backbone of this model is AE \cite{b16}. An AE can be divided into three parts input, bottleneck, and output. A bottleneck in CAE is the place where latent space is represented. 

AEs gained popularity in image compression as the basic purpose of an AE is to reduce the dimensions of the input image. Ollivier \textit{et al.} in \cite{Ollivier}, used AE to compress data by minimizing the code length. This method used the generative property of AE. In \cite{Sento}, the authors used AE with Kalman filter for the compression of the modified national institute of standards and technology (MNIST) images. The model failed to perform better for red-green-blue (RGB) images. Similarly, in \cite{Theis}, the authors used the Kodak dataset to train AE and achieved better performance than JPEG, JPEG-2000, and WebP. In \cite{Agustsson} and \cite{Dumas}, the authors used AE for the compression of images. In \cite{Dumas2}, the authors proposed AE for the image compression. They used generalized divisive normalization (GDN) and inverse generalized divisive normalization (IGDN) layers instead of rectified linear unit (ReLU) layers to boost up the training process. Cheng \textit{et al.} in \cite{Cheng} and Alexandre \textit{et al.} in \cite{Alexandre}, deployed AE for the compression of high resolution image.

\subsection{Compression with VAE}
\begin{figure*}
    \centering
    \includegraphics[width=0.6\textwidth]{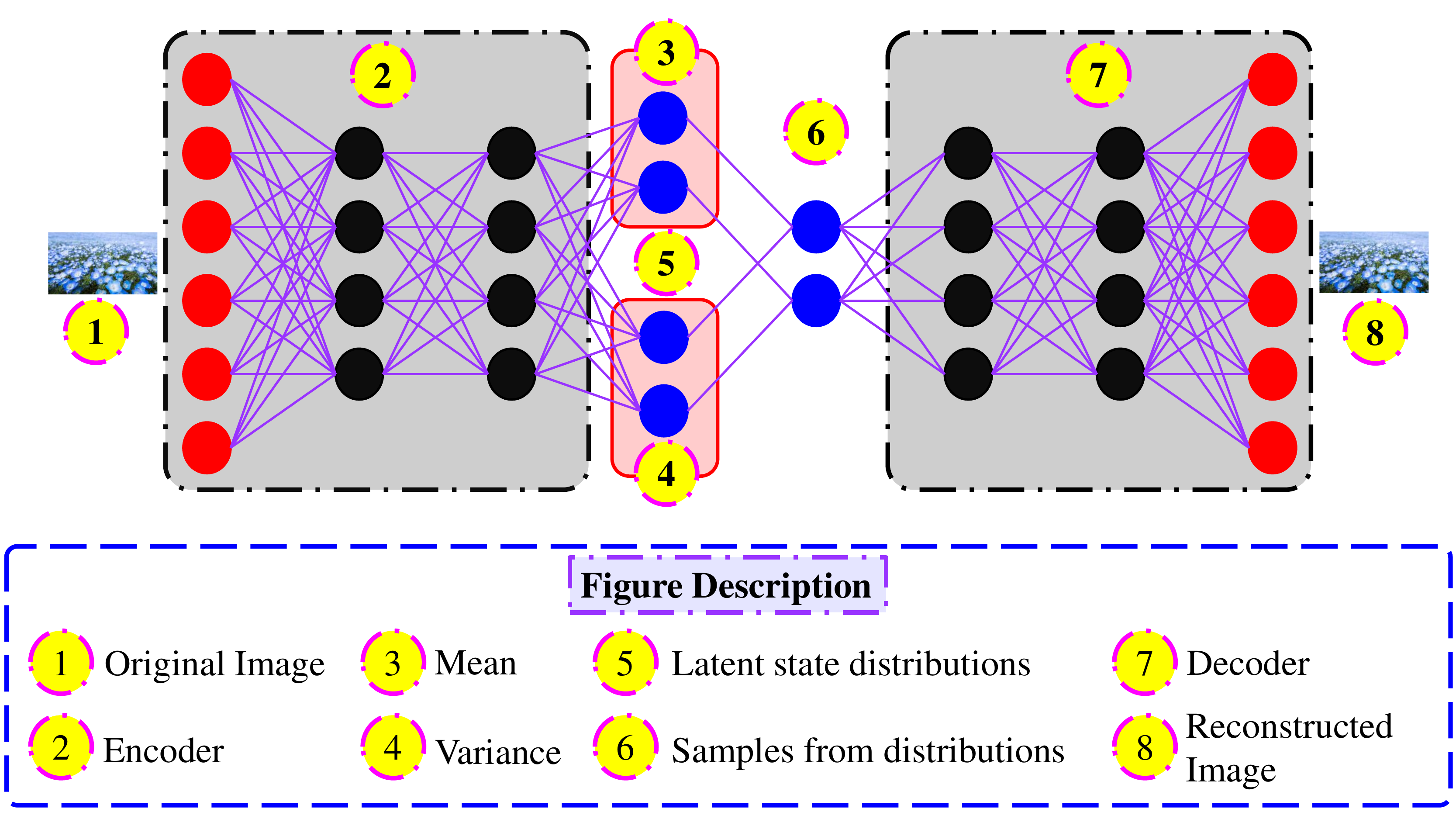}
    \caption{Image Compression with VAE.}
    \label{VAE}
\end{figure*}

VAE is a version of NN and belongs to the probabilistic graphical models and variational Bayesian methods. Compression with VAE incorporates mean and variance for latent space distributions. The results achieved by VAE are better than simple AE. A simple illustration of image compression using VAE is shown in Fig. \ref{VAE}.

Several researchers used VAE for still image compression. In \cite{Zhou}, the authors demonstrated the use of VAE having non-linear transform and uniform quantizer for image compression. Challenge one learned image compression (CLIC) dataset used for training the model. The model was complex due to massive training parameters. Recently, Chen \textit{et al.} in \cite{Chen}, proposed a VAE based architecture for high resolution image compression. The authors demonstrated a non-local attention module (NLAM) to boost the training process. However, the model complexity increased significantly. Similarly, Larsen \textit{et al.} in \cite{Larsen}, proposed a VAE to compress images. In \cite{Gregor}, the authors used VAE to compress images and achieved bits per pixel (bpp) of 4.10. However, the model was very complex due to its architecture.

\subsection{Compression with CNN}
\begin{figure*}
    \centering
    \includegraphics[width=0.6\textwidth]{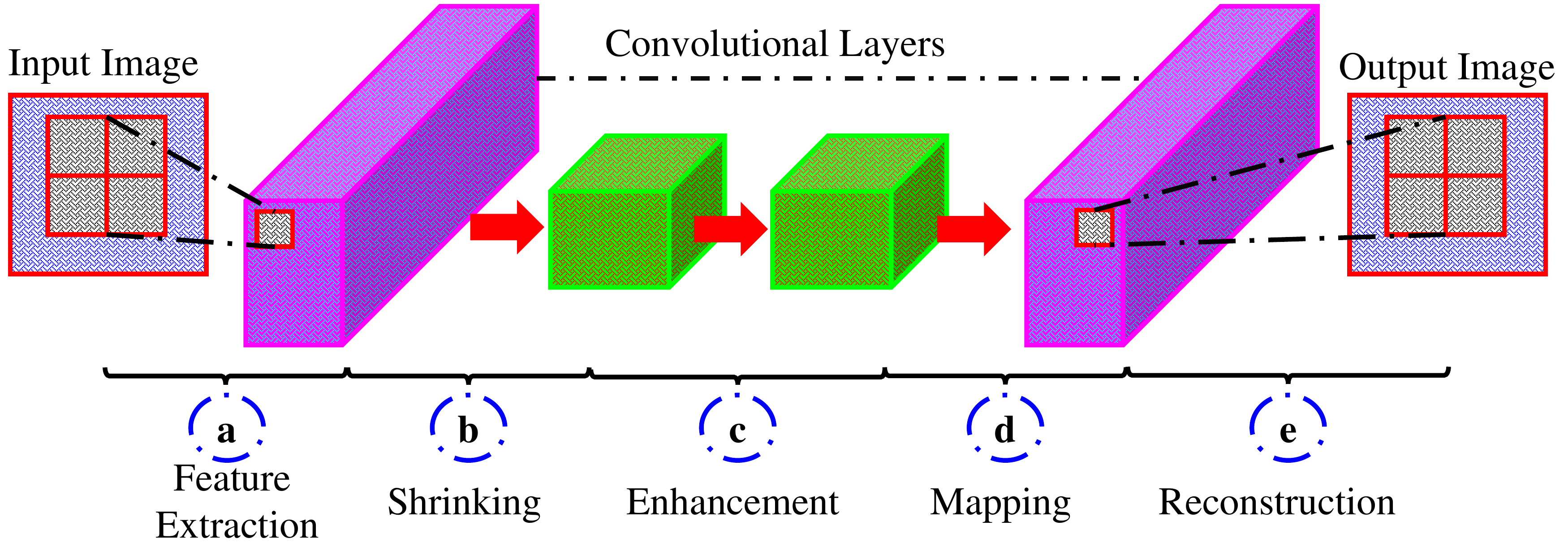}
    \caption{Image Compression and artifacts reduction with CNN.}
    \label{CNN}
\end{figure*}

CNN has prime importance in image processing due to its feature extraction characteristics. CNN is also used for image compression. The simple illustration of the image compression and artifacts reduction is shown in Fig. \ref{CNN}.

In \cite{Ayzik}--\cite{Lee}, the authors used CNNs to compress images. These CNN-based image compression models outperformed JPEG and JPEG-2000 in structural similarity index measure (SSIM) and peak signal-to-noise ratio (PSNR). In \cite{Tellez}, the authors proposed a CNN-based framework for the compression of gigapixel images. They used unsupervised learning for the training of the neural network. Similarly, in \cite{WLi}, the authors proposed a state-of-the-art optimized deep CNN-based technique for the compression of the still images. The proposed method used an attention module to compress certain regions of the images with different bits adaptively. The approach outperformed JPEG and JPEG-2000. However, the results of Ball{\'e} \cite{b15} are better. 

In \cite{Liu}, the authors also used CNN for the compression of high-resolution images. They used CLIC 2018 dataset for training and achieved 7.81\% BD-Rate reduction over BPG and JPEG-2000. In \cite{JLi}, the authors introduced a new concept of multi-spectral transform for the multi-spectral image compression using CNN. It achieved better compression efficiency than state-of-the-art anchors like JPEG-2000. Likewise, in \cite{Zeegers}, the authors proposed a data reduction CNN for the compression of hyperspectral images. Similarly, in \cite{FKong}, the authors proposed a CNN-based end-to-end compression architecture for the multi-spectral image compression. The authors extracted spatial features by partitioning and achieved better performance in terms of PSNR than JPEG-2000. Similarly, \cite{JCai2020}--\cite{Xue2019} also used CNN based frameworks. 

\subsection{Compression with RNN}
\begin{figure*}
    \centering
    \includegraphics[width=0.6\textwidth]{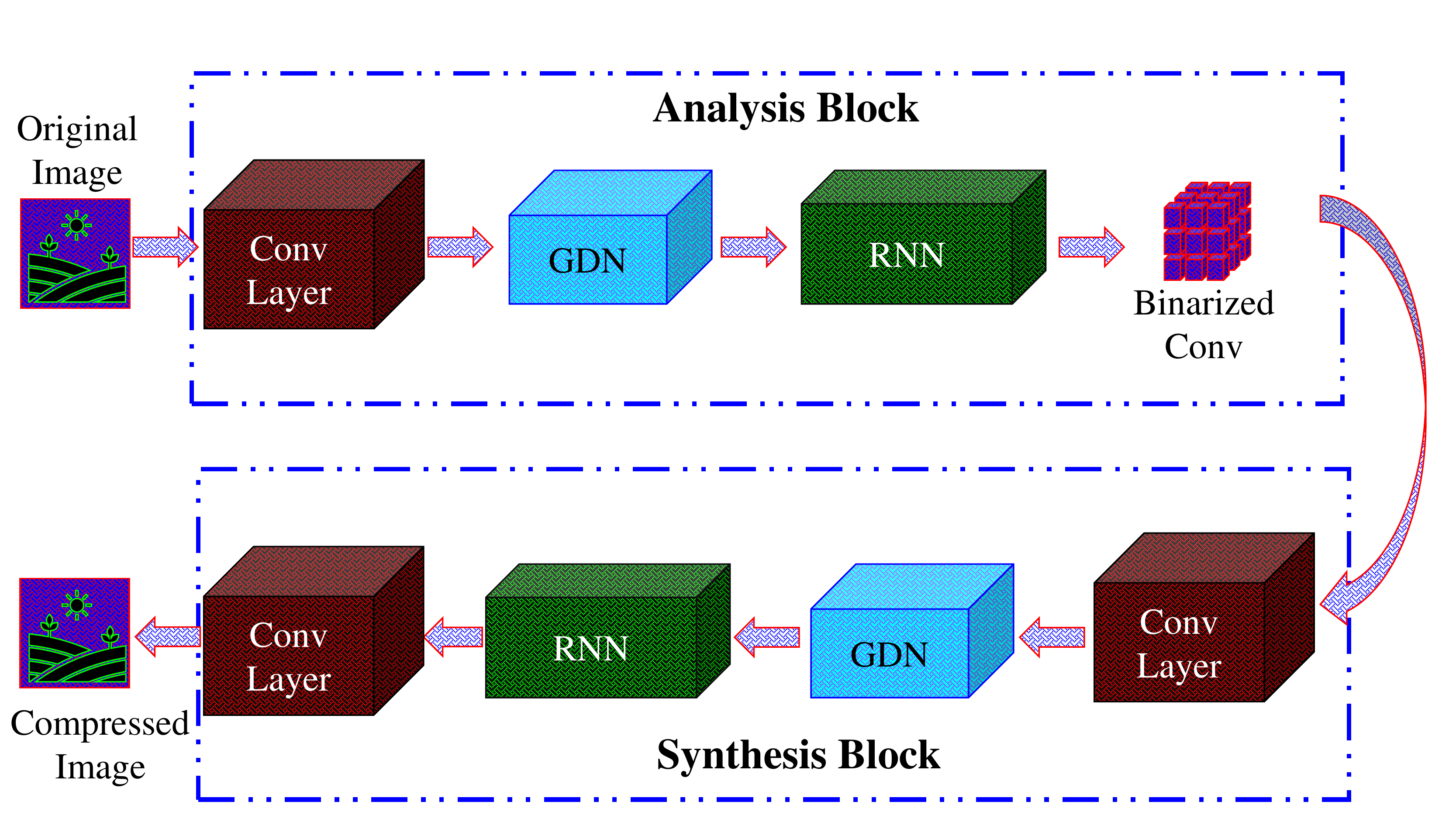}
    \caption{Image Compression with RNN.}
    \label{RNN}
\end{figure*}

RNN can be utilized for image compression too. A simple RNN-based image compression architecture has some convolutional layers, GDN layer, RNN modules, binarized convolutional layers, and IGND layer. A simple illustration image compression using RNN is shown in Fig. \ref{RNN} as proposed in \cite{Ikram}.
Covell \textit{et al.} proposed RNN based compression architecture using stop code tolerant (SCT) to train the model. They achieved bpp and PSNR of 0.25 and 27dB for Kodak and imageNet datasets, respectively. The dataset used in the method was Celeba dataset \cite{Covell}. Likewise, in \cite{Toderici}, an RNN-based method was proposed to compress still images. The training dataset used in this article was Kodak and achieved bpp of 0.5 and SSIM of 0.77. The model's performance was superior to JPEG, JPEG-2000, and WebP. 

Toderici \textit{et al.} in \cite{Toderici2}, demonstrated the use of RNN with entropy encoding to compress images. They achieved bpp of 0.5, PSNR of 33.59dB, SSIM of 0.8933, and multi-scale structural similarity index (MS-SSIM) of 0.9877. In \cite{CWang}, the authors proposed a DL-based end-to-end image compression model to compress images based on semantic analysis. The authors used RNN for the compression purpose and achieved better results than the conventional compression techniques like JPEG. The compression rate at 0.75 bpp was obtained 32 using this method. Furthermore, \cite{Punnappurath2019}-\cite{Ororbia2019} also used RNN for image compression and achieved a better compression ratio than anchors such as JPEG and JPEG-2000. 
\subsection{Compression with GAN}
\begin{figure}
    \centering
    \includegraphics[width=0.45\textwidth]{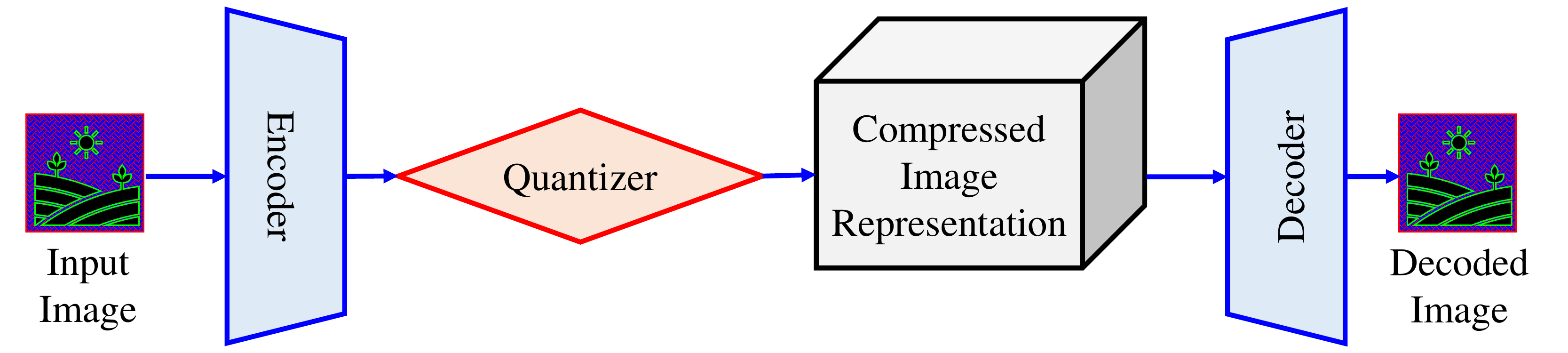}
    \caption{Image Compression with GAN.}
    \label{GAN}
\end{figure}

GANs are algorithmic architectures. GANs use two NN, pitting one against the other to generate new, synthetic data instances that can pass for actual data. These have the generator part and discriminator portion. GANs are also utilized for image compression. The simple flow diagram of image compression using GAN is shown in Fig. \ref{GAN}, where encoder and decoders are GAN modules. 

Torfason \textit{et al.} in \cite{Torfason}, illustrated the use of GAN for the compression and classification of semantic data. Similarly, \cite{EAgustsson} proposed an extreme image compression architecture using GAN. They used two types of modules for image compression. The first module was generative compression, where no semantic label maps were required. The type applied was selective generative compression, where semantic label maps were used. They provided insight towards the full resolution image compression and targeted low bit-rate, i.e., less than 0.1 bpp. In \cite{JSong}, the authors demonstrated the use of unified binary GAN (BGAN+) for image compression and image retrieval. The model achieved better than JPEG and JPEG-2000. The visual quality of the reconstructed image was much improved than JPEG and JPEG-2000 at low bit-rate such as 0.15 bpp. In \cite{Zhang2018}-\cite{Galteri2017}, the authors proposed GAN-based architectures for the image compression and achieved high compression efficiency. The drawback of the GAN-based architecture is the cost of deployment.
\subsection{Compression with PCA}
PCA is also used for image compression. In \cite{Bascones}, the authors utilized vector quantization and PCA for the compression of hyperspectral images. The technique achieved better performance than JPEG-2000. Similarly, in \cite{Yadav2017}, the authors used PCA and DCT to compress hyperspectral images. The used PSNR metric for the evaluation of the quality of the image. The method applies to hyperspectral images only.

In \cite{Karaca2018}, the authors used weighted PCA and JPEG-2000 to compress hyperspectral images. They demonstrated that the hybrid approach of weighted PCA and JPEG-2000 achieved better PSNR than the JPEG and JPEG-2000. Similarly, the authors of \cite{Abbas2018} proposed PCA for lossy compression and achieved quality reconstructed image. However, the method depended on the number of components, and the performance decreased with the increase of PCA components.  
\subsection{Compression with fuzzy means clustering}
Martino \textit{et al.}, in \cite{Martino1} proposed a fast algorithm to improve the performance of the multilevel fuzzy transform image compression method \cite{Martino2}. The authors demonstrated that the reconstructed image's visual quality.

Moreover, in \cite{WKhalaf} the authors proposed a conventional technique of using a curve-fitting novel hyperbolic function for the compression. The main advantages of this approach were: strengthening the edges of the image, removing the blocking effect, improving the SSIM, and increasing the PSNR up to 20 dB.  
\begin{table*}
\centering
\caption{Comparison of the learning-driven lossy image compression models.}
\begin{tabular}{p{45pt} p{45pt} p{45pt} p{30pt} p{30pt} p{30pt} p{200pt}}
    \hline
    \hline
     \vspace{0.01cm}\textbf{Research} \vspace{0.1cm}&\vspace{0.01cm} \textbf{Technique} \vspace{0.1cm}&\vspace{0.01cm} \textbf{bpp} \vspace{0.1cm}& \multicolumn{3}{p{90pt}}{\vspace{0.01cm}\textbf{Performance Parameters}\vspace{0.1cm}} & \vspace{0.01cm}\textbf{Contributions and limitations}\vspace{0.1cm}\\\cline{4-6}
     \vspace{0.01cm} \vspace{0.1cm}& \vspace{0.01cm} \vspace{0.1cm}&\vspace{0.01cm} \vspace{0.1cm}&\vspace{0.01cm} \textbf{PSNR [dB]} \vspace{0.1cm}&\vspace{0.01cm} \textbf{SSIM} \vspace{0.1cm}& \vspace{0.01cm}\textbf{MS-SSIM}\vspace{0.1cm} &\vspace{0.01cm} \vspace{0.1cm}\\
     \hline
     \hline
    \vspace{0.01cm}\cite{Ollivier}\vspace{0.1cm}& 
    \vspace{0.01cm}AE \vspace{0.1cm}&
    \vspace{0.01cm}-- \vspace{0.1cm}& 
    \vspace{0.01cm}-- \vspace{0.1cm}& 
    \vspace{0.01cm}-- \vspace{0.1cm}& 
    \vspace{0.01cm}-- \vspace{0.1cm}& 
    \vspace{0.01cm}Generative property of AEs, by minimizing code-length, limited to theoretical perspective.\vspace{0.1cm}\\
    \hline
    \vspace{0.01cm}\cite{Sento}\vspace{0.1cm}&
    \vspace{0.01cm}AE \vspace{0.1cm}&
    \vspace{0.01cm}-- \vspace{0.1cm}& 
    \vspace{0.01cm}-- \vspace{0.1cm}& 
    \vspace{0.01cm}-- \vspace{0.1cm}& 
    \vspace{0.01cm}-- \vspace{0.1cm}& 
    \vspace{0.01cm} The combination of the non-recurrent architecture of three-layers with kalman filter, limited to grayscale images.
    \vspace{0.1cm}\\
    \hline
    \vspace{0.01cm}\cite{Theis}\vspace{0.1cm}&
    \vspace{0.01cm}AE \vspace{0.1cm}&
    \vspace{0.01cm}0.4 \vspace{0.1cm}& 
    \vspace{0.01cm}29 \vspace{0.1cm}&
    \vspace{0.01cm}0.83 \vspace{0.1cm}& 
    \vspace{0.01cm}0.94 \vspace{0.1cm}& 
    \vspace{0.01cm} Compressive AE with MSE loss function trained by residual network and non-differentiability of the quantization noise reduced using rounding based quantization, visual artifacts problem at low bit rate.  \vspace{0.1cm}\\
    \hline
    \vspace{0.01cm}\cite{Agustsson}\vspace{0.1cm}& 
    \vspace{0.01cm}AE \vspace{0.1cm}&
    \vspace{0.01cm}0.2 \vspace{0.1cm}& 
    \vspace{0.01cm}--\vspace{0.1cm}&
    \vspace{0.01cm}-- \vspace{0.1cm}& 
    \vspace{0.01cm}0.92 \vspace{0.1cm}& 
    \vspace{0.01cm}Vector quantization preferred over scalar quantization, joint optimization of learning of latent space representation, worst performance than BPG.\vspace{0.1cm}\\
    \hline
    \vspace{0.01cm}\cite{Dumas}\vspace{0.1cm}&
    \vspace{0.01cm}AE \vspace{0.1cm}&
    \vspace{0.01cm}0.1 \vspace{0.1cm}& 
    \vspace{0.01cm}31.5 \vspace{0.1cm}&
    \vspace{0.01cm}-- \vspace{0.1cm}& 
    \vspace{0.01cm}-- \vspace{0.1cm}& 
    \vspace{0.01cm}Global rate-distortion (RD) constraint in a framework stochastic winner-take-all autoencoder (SWTA-AE), worst performance than JPEG-2000, changing strides of convolutional layers harms NN.\vspace{0.1cm}\\
    \hline
    \vspace{0.01cm}\cite{Dumas2}\vspace{0.1cm}& 
    \vspace{0.01cm}AE\vspace{0.1cm}& 
    \vspace{0.01cm}0.2\vspace{0.1cm}&
    \vspace{0.01cm}31 \vspace{0.1cm}& 
    \vspace{0.01cm}-- \vspace{0.1cm}& 
    \vspace{0.01cm}-- \vspace{0.1cm}& \vspace{0.01cm}GDN as well as IGDN layers, worst performance than HEVC intra.
    \vspace{0.1cm}\\
    \hline
    \vspace{0.01cm}\cite{Cheng}\vspace{0.1cm}&
    \vspace{0.01cm}AE\vspace{0.1cm}& 
    \vspace{0.01cm}1.2 \vspace{0.1cm}& 
    \vspace{0.01cm}33 \vspace{0.1cm}& 
    \vspace{0.01cm}-- \vspace{0.1cm}& 
    \vspace{0.01cm}-- \vspace{0.1cm}& 
    \vspace{0.01cm}Parametric ReLU activation function, to produce feature maps with low dimensionality, uses a CAE instead of a transform and an inverse transform, along with some max-pooling-up-sampling layers, comparable performance with Ball{'\ e} model for grayscale images only.\vspace{0.1cm}\\
    \hline
    \vspace{0.01cm}\cite{Alexandre}\vspace{0.1cm}& 
    \vspace{0.01cm}AE \vspace{0.1cm}&
    \vspace{0.01cm}0.126\vspace{0.1cm}& 
    \vspace{0.01cm}29.30 \vspace{0.1cm}&
    \vspace{0.01cm}-- \vspace{0.1cm}& 
    \vspace{0.01cm}0.924 \vspace{0.1cm}& 
     \vspace{0.01cm}Feature map coding block with Importance Net, Uses skip connections AEs and trained with loss function of MSE and MS-SSIM, less PSNR than BPG.
    \vspace{0.1cm}\\
     \hline
     \vspace{0.01cm}\cite{Zhou}\vspace{0.1cm}& 
     \vspace{0.01cm}VAE \vspace{0.1cm}& 
    \vspace{0.01cm}0.15 \vspace{0.1cm}& 
     \vspace{0.01cm}30.76 \vspace{0.1cm}&
     \vspace{0.01cm}-- \vspace{0.1cm}& 
     \vspace{0.01cm}0.955 \vspace{0.1cm}& 
     \vspace{0.01cm}Non-linear encoder (NLE) transform, uniform quantizer, non-linear decoder (NLD) transform, and post-processing module form a VAE framework of article. \vspace{0.1cm}\\
    \hline
    \vspace{0.01cm}\cite{Chen}\vspace{0.1cm}& 
     \vspace{0.01cm}VAE \vspace{0.1cm}&
     \vspace{0.01cm}0.2 \vspace{0.1cm}&
     \vspace{0.01cm}30 \vspace{0.1cm}&
     \vspace{0.01cm}-- \vspace{0.1cm}& 
     \vspace{0.01cm}0.7768 \vspace{0.1cm}& 
     \vspace{0.01cm}Non local (NL) operations to consider local as well as global context, better performance than Ball{'\ e} model.\vspace{0.1cm}\\
    \hline
     \vspace{0.01cm}\cite{Larsen}\vspace{0.1cm}&      \vspace{0.01cm}VAE \vspace{0.1cm}& 
     \vspace{0.01cm}-- \vspace{0.1cm}& 
    \vspace{0.01cm}-- \vspace{0.1cm}&
    \vspace{0.01cm}-- \vspace{0.1cm}& 
    \vspace{0.01cm}-- \vspace{0.1cm}& 
     \vspace{0.01cm}Kullback-Leiber (KL) loss function, limited to image generation.
     \vspace{0.1cm}\\
     \hline
    \vspace{0.01cm}\cite{Gregor}\vspace{0.1cm}& 
    \vspace{0.01cm}VAE \vspace{0.1cm}& 
    \vspace{0.01cm}4.10 \vspace{0.1cm}& 
    \vspace{0.01cm}--\vspace{0.1cm}&
    \vspace{0.01cm}-- \vspace{0.1cm}& 
    \vspace{0.01cm}-- \vspace{0.1cm}& 
    \vspace{0.01cm}Recurrent VAE for learning latent space, limited to conceptual compression.
    \vspace{0.1cm}\\
    \hline
    \vspace{0.01cm}\cite{Ayzik}\vspace{0.1cm}&
    \vspace{0.01cm}CNN \vspace{0.1cm}& 
    \vspace{0.01cm}0.025 \vspace{0.1cm}&
    \vspace{0.01cm}-- \vspace{0.1cm}& 
    \vspace{0.01cm}-- \vspace{0.1cm}& 
    \vspace{0.01cm}0.925 \vspace{0.01cm}& 
    \vspace{0.01cm} Initial decoded image and side information require more complexity and storage space, cloud based model, visual artifacts at low bpp.\vspace{0.1cm}\\
    \hline
    \vspace{0.01cm}\cite{Hu}\vspace{0.1cm}& 
    \vspace{0.01cm}CNN \vspace{0.1cm}& 
    \vspace{0.01cm}0.25 \vspace{0.1cm}& 
    \vspace{0.01cm}30 \vspace{0.1cm}&
    \vspace{0.01cm}-- \vspace{0.1cm}& 
    \vspace{0.01cm}-- \vspace{0.1cm}& 
    \vspace{0.01cm}Coarse to fine hyperprior based model, lacks parallel decoding.\vspace{0.1cm}\\
    \hline
    \vspace{0.01cm}\cite{Raman}\vspace{0.1cm}&
    \vspace{0.01cm}CNN \vspace{0.1cm}& 
    \vspace{0.01cm}0.0726 \vspace{0.1cm}& 
    \vspace{0.01cm}23.93\vspace{0.1cm}&
    \vspace{0.01cm}0.8118 \vspace{0.1cm}& 
    \vspace{0.01cm}-- \vspace{0.1cm}& 
    \vspace{0.01cm}Loss function consisting of MSE, adversarial and layer wise loss.
    \vspace{0.1cm}\\
    \hline
    \vspace{0.01cm}\cite{Cheng2}\vspace{0.1cm}& 
    \vspace{0.01cm}CNN \vspace{0.1cm}&
    \vspace{0.01cm}0.519 \vspace{0.1cm}& 
    \vspace{0.01cm}33.62 \vspace{0.1cm}& 
    \vspace{0.01cm}-- \vspace{0.1cm}& 
    \vspace{0.01cm}0.981 \vspace{0.1cm}& 
    \vspace{0.01cm}The Lagrange multiplier for joint rate-distortion, considerable run-time complexity due to the attention modules.\vspace{0.1cm}\\
    \hline
    \vspace{0.01cm}\cite{Lee}\vspace{0.1cm}& 
    \vspace{0.01cm}CNN \vspace{0.1cm}& 
    \vspace{0.01cm}0.2 \vspace{0.1cm}& 
    \vspace{0.01cm}31 \vspace{0.1cm}&
    \vspace{0.01cm}-- \vspace{0.1cm}& 
    \vspace{0.01cm}0.7878 \vspace{0.1cm}& 
    \vspace{0.01cm}Gaussian mixture module (GMM) alogwith JointIQ-Net to outperform JPEG-2k, BPG, versatile video coding (VVC) Intra scheme.
    \vspace{0.1cm}\\
    \hline
    \vspace{0.01cm}\cite{Tellez}\vspace{0.1cm}& 
    \vspace{0.01cm}CNN\vspace{0.1cm}& 
    \vspace{0.01cm}--\vspace{0.1cm}&
    \vspace{0.01cm}--\vspace{0.1cm}&
    \vspace{0.01cm}-- \vspace{0.1cm}& 
    \vspace{0.01cm}-- \vspace{0.1cm}& 
    \vspace{0.01cm}Two step CNN based framework and the unsupervised training, limited to histopathology images.\vspace{0.1cm}\\
    \hline
    \vspace{0.01cm}\cite{WLi}\vspace{0.1cm}& 
         \vspace{0.01cm}CNN \vspace{0.1cm}& 
         \vspace{0.01cm}0.5 \vspace{0.1cm}& 
         \vspace{0.01cm}-- \vspace{0.1cm}& 
         \vspace{0.01cm}0.77\vspace{0.1cm}&
         \vspace{0.01cm}-- \vspace{0.1cm}& 
         \vspace{0.01cm}Better performance than JPEG and JPEG-2k but worst performance than Ball{\'e}. \cite{b15} \vspace{0.1cm}\\
         \hline
    \hline
\end{tabular}
\label{Table1}
\end{table*}

\begin{table*}
    \centering
     \caption{Comparison of the learning-driven lossy image compression models.}
    \begin{tabular}{p{45pt} p{45pt} p{45pt} p{30pt} p{30pt} p{30pt} p{200pt}}
    \hline
    \hline
     \vspace{0.01cm}\textbf{Research} \vspace{0.1cm}&\vspace{0.01cm} \textbf{Technique} \vspace{0.1cm}&\vspace{0.01cm} \textbf{bpp} \vspace{0.1cm}& \multicolumn{3}{p{90pt}}{\vspace{0.01cm}\textbf{Performance Parameters}\vspace{0.1cm}} & \vspace{0.01cm}\textbf{Contributions and limitations}\vspace{0.1cm}\\\cline{4-6}
     \vspace{0.01cm} \vspace{0.1cm}& \vspace{0.01cm} \vspace{0.1cm}&\vspace{0.01cm}  \vspace{0.1cm}&\vspace{0.01cm} \textbf{PSNR [dB]} \vspace{0.1cm}&\vspace{0.01cm} \textbf{SSIM} \vspace{0.1cm}& \vspace{0.01cm}\textbf{MS-SSIM}\vspace{0.1cm} &\vspace{0.01cm} \vspace{0.1cm}\\
     \hline
     \hline
              \vspace{0.01cm}\cite{Liu}\vspace{0.1cm}& 
         \vspace{0.01cm}CNN \vspace{0.1cm}& 
         \vspace{0.01cm}0.2 \vspace{0.1cm}& 
         \vspace{0.01cm}-- \vspace{0.1cm}& 
         \vspace{0.01cm}-- \vspace{0.1cm}& 
         \vspace{0.01cm}0.93\vspace{0.1cm}&
         \vspace{0.01cm}Trained with perceptual as well as adversarial loss for the generation of sharp details, outperforms JPEG, JPEG-2000, BPG and WebP.\vspace{0.1cm}\\
         \hline
         \vspace{0.01cm}\cite{JLi}\vspace{0.1cm}& 
         \vspace{0.01cm}CNN \vspace{0.1cm}& 
         \vspace{0.01cm}-- \vspace{0.1cm}& 
         \vspace{0.01cm}--\vspace{0.1cm}&
         \vspace{0.01cm}-- \vspace{0.1cm}& 
         \vspace{0.01cm}-- \vspace{0.1cm}& 
         \vspace{0.01cm}Multi-spectral transform for the multi-spectral image compression.\vspace{0.1cm}\\
         \hline
         \vspace{0.01cm}\cite{Zeegers}\vspace{0.1cm}& 
         \vspace{0.01cm}CNN \vspace{0.1cm}& 
         \vspace{0.01cm}--\vspace{0.1cm}&
         \vspace{0.01cm}--\vspace{0.1cm}&
         \vspace{0.01cm}-- \vspace{0.1cm}& 
         \vspace{0.01cm}-- \vspace{0.1cm}& 
         \vspace{0.01cm}Introduction of data reduction block to save GPU memory.
         \vspace{0.1cm}\\
         \hline
         \vspace{0.01cm}\cite{FKong}\vspace{0.1cm}&
         \vspace{0.01cm}CNN \vspace{0.1cm}& 
         \vspace{0.01cm}-- \vspace{0.1cm}& 
         \vspace{0.01cm}--\vspace{0.1cm}&
         \vspace{0.01cm}-- \vspace{0.1cm}& 
         \vspace{0.01cm}-- \vspace{0.1cm}& 
         \vspace{0.01cm}Feature extraction by partitioning, better performance than JPEG-2000.
         \vspace{0.1cm}\\
         \hline
         \vspace{0.01cm}\cite{Covell}\vspace{0.1cm}& 
         \vspace{0.01cm}RNN  \vspace{0.1cm}& 
         \vspace{0.01cm}0.25 \vspace{0.1cm}& 
         \vspace{0.01cm}27 \vspace{0.1cm}&
         \vspace{0.01cm}-- \vspace{0.1cm}& 
         \vspace{0.01cm}-- \vspace{0.1cm}& 
         \vspace{0.01cm}RNN trained using SCT, limited to adaptive encoding. \vspace{0.1cm}\\
         \hline
         \vspace{0.01cm}\cite{Toderici}\vspace{0.1cm}& 
         \vspace{0.01cm}RNN \vspace{0.1cm}& 
         \vspace{0.01cm}0.5 \vspace{0.1cm}&
         \vspace{0.01cm}-- \vspace{0.1cm}& 
         \vspace{0.01cm}0.77\vspace{0.1cm}&
         \vspace{0.01cm}-- \vspace{0.1cm}& 
         \vspace{0.01cm}Focuses on 32x32 size thumbnail images, better than JPEG, JPEG-2000 and WebP. \vspace{0.1cm}\\
         \hline
         \vspace{0.01cm}\cite{Toderici2}\vspace{0.1cm}& 
         \vspace{0.01cm}RNN\vspace{0.1cm} & 
         \vspace{0.01cm}0.5 \vspace{0.1cm}& 
         \vspace{0.01cm}33.59 \vspace{0.1cm}&
         \vspace{0.01cm}0.8933 \vspace{0.1cm}& 
         \vspace{0.01cm}0.9877 \vspace{0.1cm}& 
         \vspace{0.01cm}Long short-term memory (LSTM)-based progressive encoding of thumbnail pictures $(32 \times 32)$, decent compression performance at low bit rates, pixel RNN for entropy coding.\vspace{0.1cm}\\
         \hline
         \vspace{0.01cm}\cite{CWang}\vspace{0.1cm}&
         \vspace{0.01cm}RNN \vspace{0.1cm}& 
         \vspace{0.01cm}0.75 \vspace{0.1cm}&
         \vspace{0.01cm}32 \vspace{0.1cm}& 
         \vspace{0.01cm}-- \vspace{0.1cm}& 
         \vspace{0.01cm}-- \vspace{0.1cm}&
         \vspace{0.01cm}Semantic analysis, better performance than JPEG. \vspace{0.1cm}\\
         \hline
         \vspace{0.01cm}\cite{Torfason}\vspace{0.1cm}& 
         \vspace{0.01cm}GAN \vspace{0.1cm}& 
         \vspace{0.01cm}0.0983 \vspace{0.1cm}& 
         \vspace{0.01cm}28.54 \vspace{0.1cm}& 
         \vspace{0.01cm}0.85 \vspace{0.1cm}& 
         \vspace{0.01cm}0.973 \vspace{0.1cm}& 
         \vspace{0.01cm}For semantic understanding combined training for compression and classification. \vspace{0.1cm}\\
         \hline
         \vspace{0.01cm}\cite{EAgustsson}\vspace{0.1cm}& 
         \vspace{0.01cm}GAN \vspace{0.1cm}& 
         \vspace{0.01cm}0.033 \vspace{0.1cm}& 
         \vspace{0.01cm}--\vspace{0.1cm}&
         \vspace{0.01cm}-- \vspace{0.1cm}& 
         \vspace{0.01cm}-- \vspace{0.1cm}& 
         \vspace{0.01cm}Full-resolution image compression, applicable only when semantic labels are available. \vspace{0.1cm}\\
         \hline
         \vspace{0.01cm}\cite{JSong}\vspace{0.1cm}& 
         \vspace{0.01cm}BGAN+ \vspace{0.1cm}& 
         \vspace{0.01cm}0.15 \vspace{0.1cm}& 
         \vspace{0.01cm}-- \vspace{0.1cm}& 
         \vspace{0.01cm}-- \vspace{0.1cm}& 
         \vspace{0.01cm}0.95\vspace{0.1cm}& 
         \vspace{0.01cm}Reconstructed image without distorting the small objects present in the image at low bit-rates such as 0.15bpp. \vspace{0.1cm}\\
         \hline
         \vspace{0.01cm}\cite{Bascones}\vspace{0.1cm}& 
         \vspace{0.01cm}PCA \vspace{0.1cm}& 
         \vspace{0.01cm}-- \vspace{0.1cm}& 
         \vspace{0.01cm}-- \vspace{0.1cm}&
         \vspace{0.01cm}-- \vspace{0.1cm}& 
         \vspace{0.01cm}-- \vspace{0.1cm}& 
         \vspace{0.01cm}Better performance than JPEG-2000, limited to hyper-spectral data. \vspace{0.1cm}\\
         \hline
         \vspace{0.01cm}\cite{Martino1}-\cite{Martino2}\vspace{0.1cm}& 
         \vspace{0.01cm}Multilevel Fuzzy Transform \vspace{0.1cm}& 
         \vspace{0.01cm}-- \vspace{0.1cm}& 
         \vspace{0.01cm}-- \vspace{0.1cm}&
         \vspace{0.01cm}-- \vspace{0.1cm}& 
         \vspace{0.01cm}-- \vspace{0.1cm}& 
         \vspace{0.01cm}Reconstructed image with better visual quality, limited to maximum of 1024$\times$1024 image resolutions. \vspace{0.1cm}\\
         \hline
         \vspace{0.01cm}\cite{WKhalaf}\vspace{0.1cm}& 
         \vspace{0.01cm}Curve-fitting with hyperbolic function \vspace{0.1cm}& 
         \vspace{0.01cm}-- \vspace{0.1cm}& 
         \vspace{0.01cm}20 \vspace{0.1cm}&
         \vspace{0.01cm}-- \vspace{0.1cm}& 
         \vspace{0.01cm}-- \vspace{0.1cm}& 
         \vspace{0.01cm}Removed blocking effects and improved SSIM, results are dependent on block size and compression ratio. \vspace{0.1cm}\\
         \hline
         \hline
    \end{tabular}
    \label{Table2}
\end{table*}

\begin{figure}[t]
    \centering
    \includegraphics[width=0.5\textwidth]{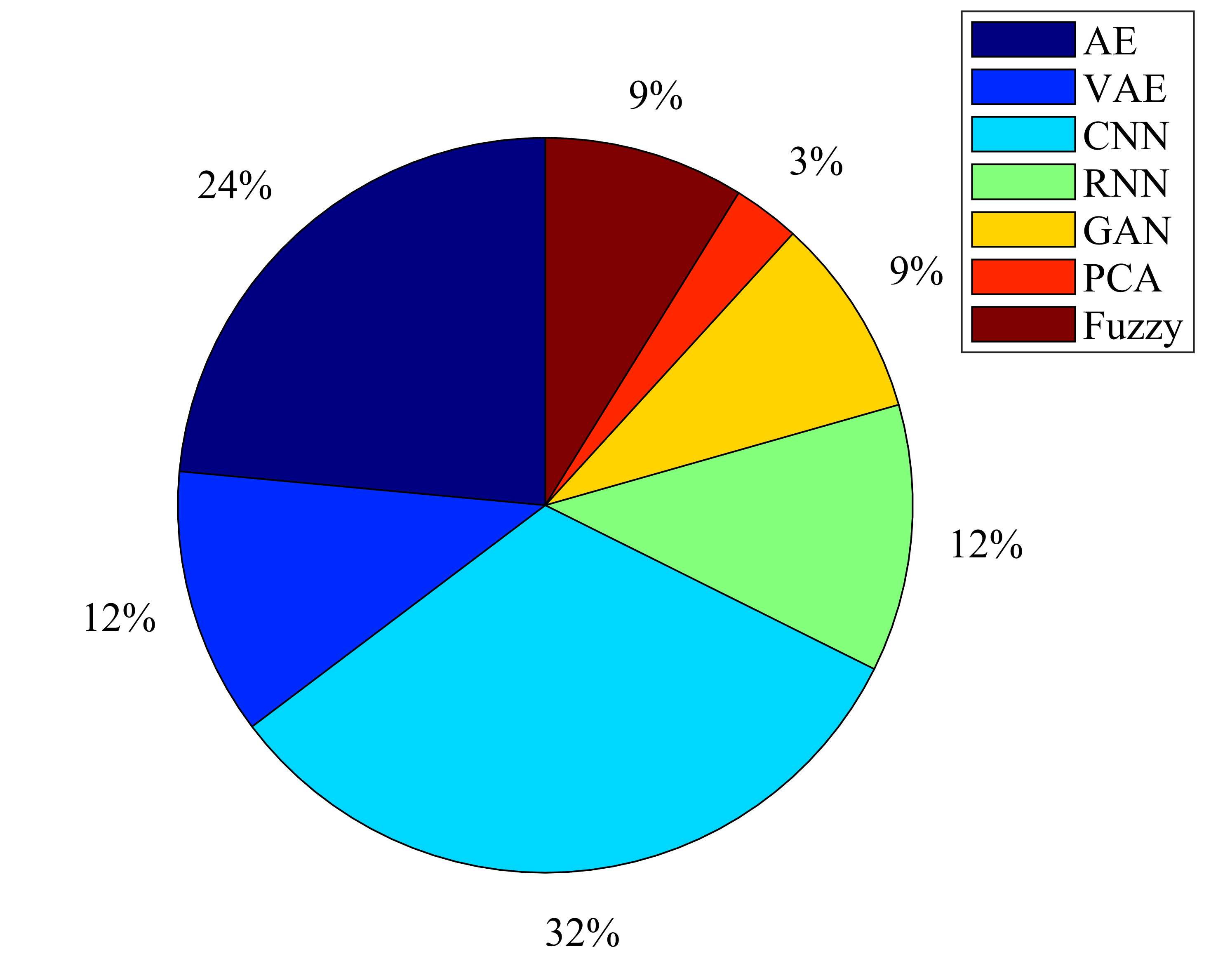}
    \caption{Percentage of architectures present for still image compression.}
    \label{Comp}
\end{figure}

Table \ref{Table1} and \ref{Table2} present a detailed summary of the techniques used for image compression. Fig. \ref{Comp} shows the percentage of the different ML architectures present in literature for still image compression. It is evident that 24\% of architectures in the literature are based on AEs while 12\% are based on VAEs. The majority of the architectures are based on CNN 32\%. 
\subsection{Comparative Analysis of State-of-the-art Techniques}
\begin{figure*}
    \centering
    \includegraphics[width=0.8\textwidth]{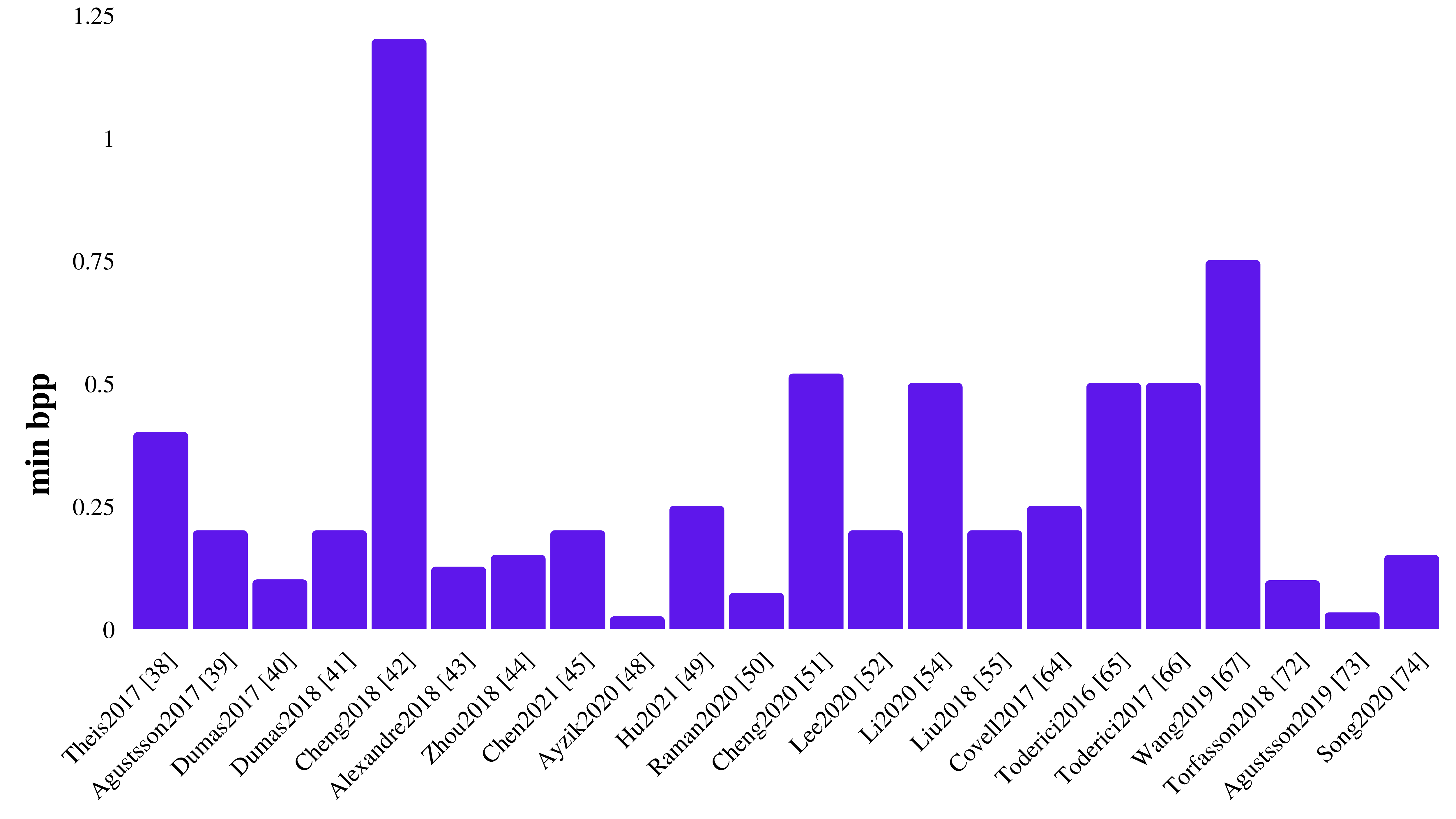}
    \caption{Minimum bpp achieved by different models.}
    \label{bpp}
\end{figure*}

\begin{figure*}
    \centering
    \includegraphics[width=0.8\textwidth]{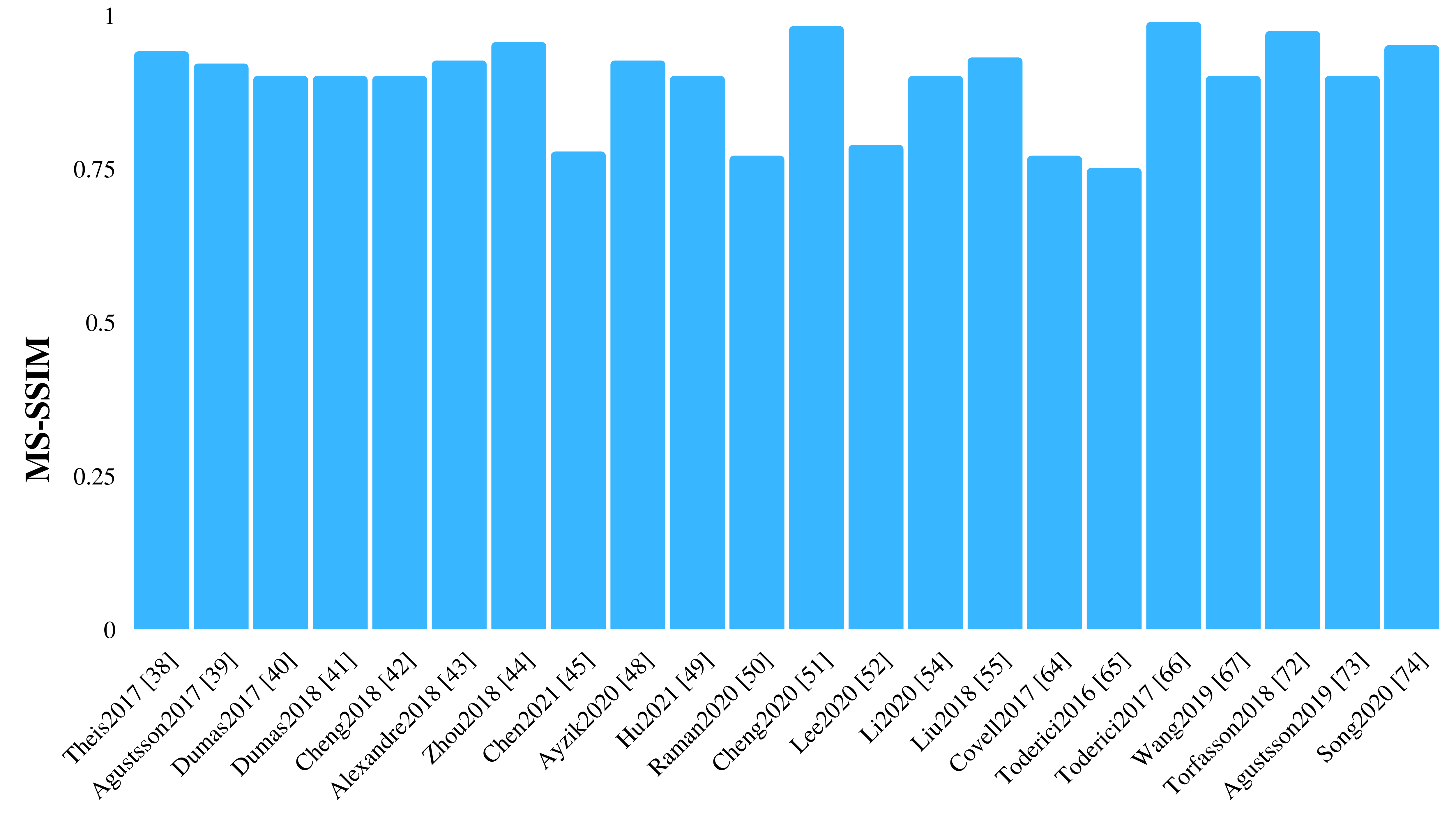}
    \caption{MS-SSIM achieved at minimum bpp by different models.}
    \label{msssim}
\end{figure*}

\begin{figure*}
    \centering
    \includegraphics[width=0.8\textwidth]{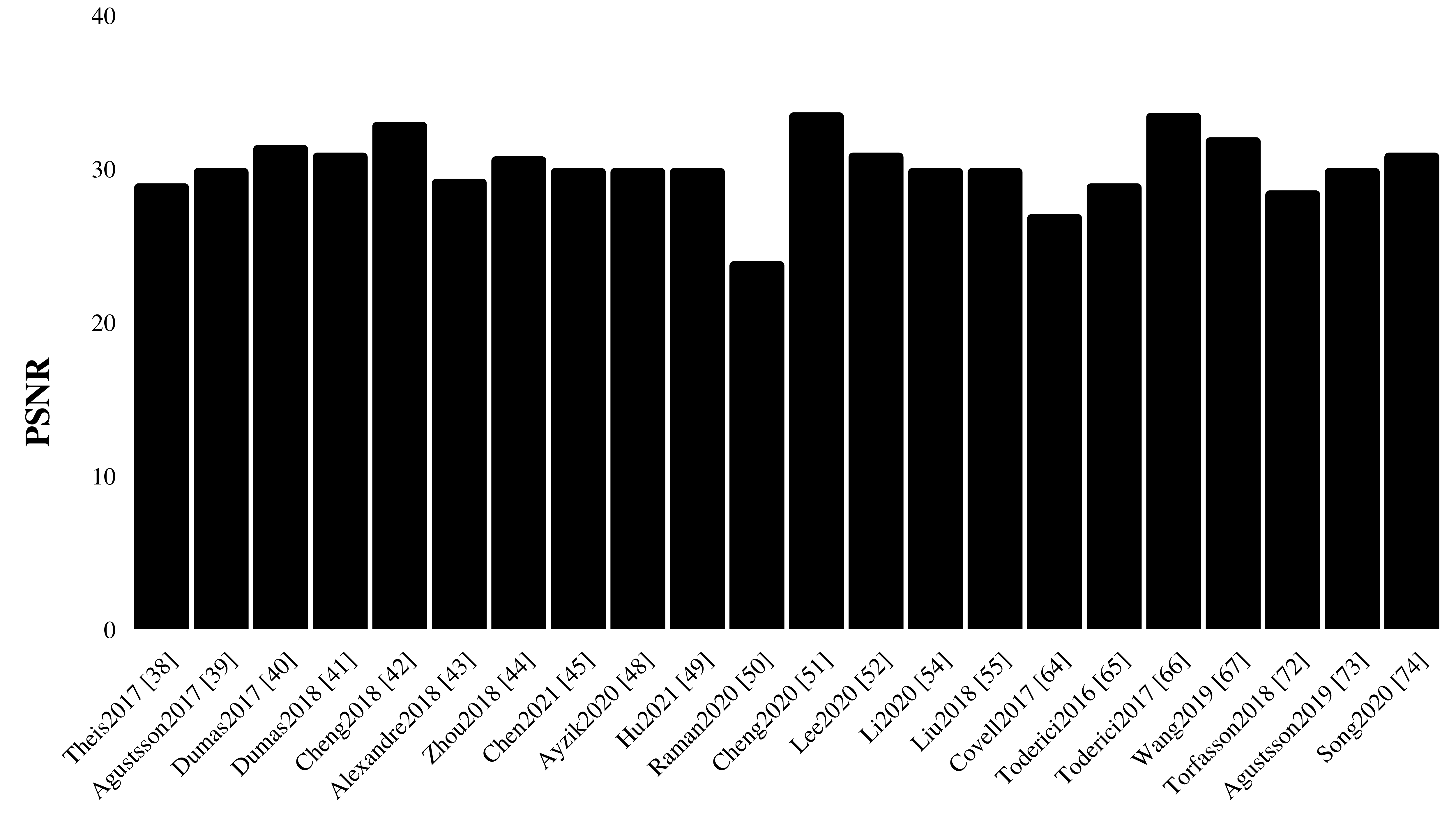}
    \caption{PSNR achieved at minimum bpp by different models.}
    \label{psnr}
\end{figure*}

\begin{figure}[t]
    \centering
    \includegraphics[width=0.45\textwidth]{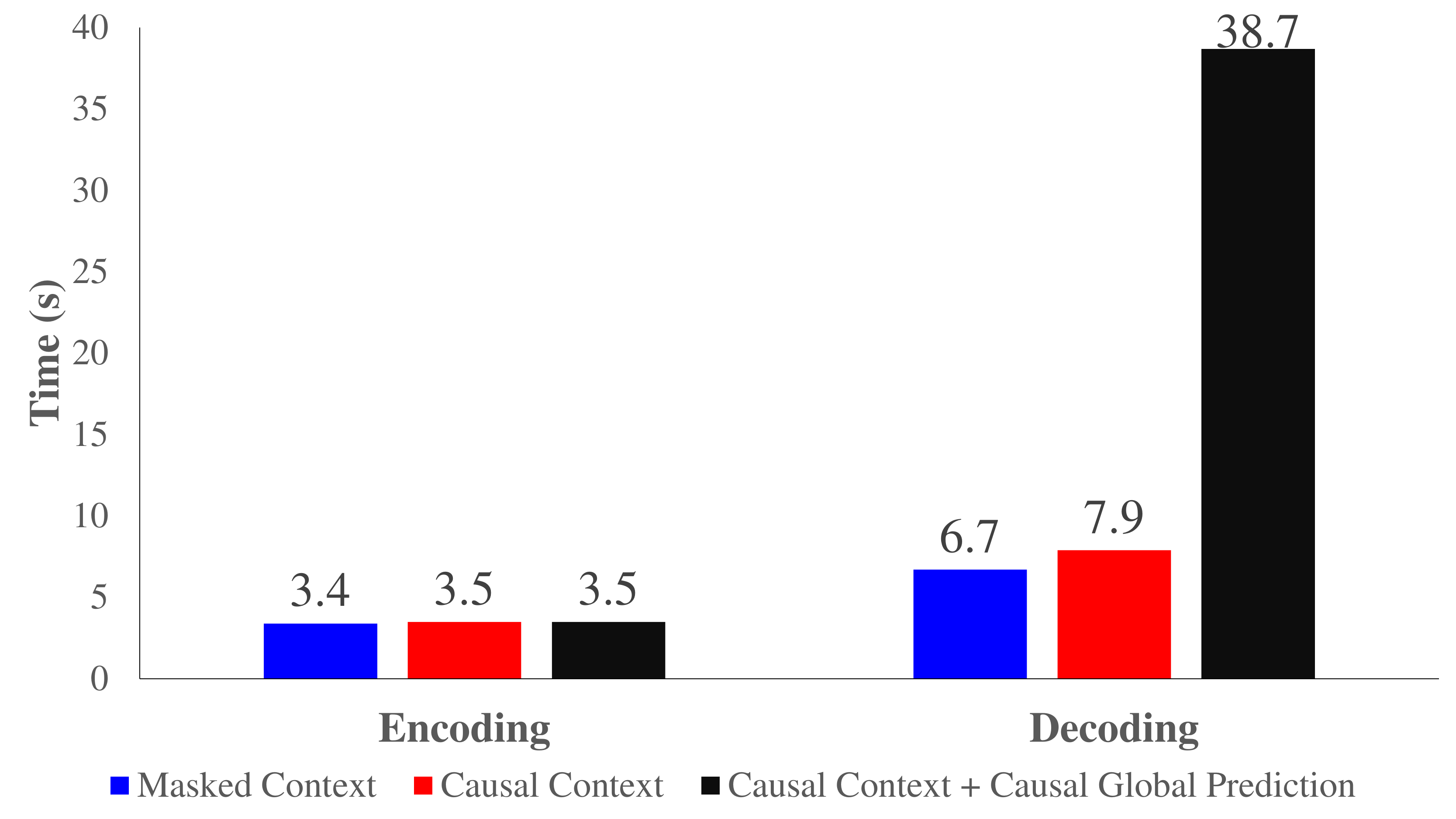}
    \caption{Comparison of encoding and decoding time of masked context, causal context and causal context plus causal global prediction.}
    \label{EncodingTime}
\end{figure}

This subsection presents a comparative analysis of state-of-the-art techniques in terms of compression ratio, computational time, PSNR, MS-SSIM, bpp, and RD-rate. Fig. \ref{bpp} shows the minimum bpp of the learning-driven lossy image compression models. Fig. \ref{msssim} and Fig. \ref{psnr} show the corresponding MS-SSIM and PSNR achieved at minimum bpp by reference respectively. Ayzik \textit{et al.} \cite{Ayzik} achieved PSNR of 30dB and MS-SSIM of 0.925 at the minimum bpp of 0.025. The comparison of the three context models as presented in \cite{ZGuo} shows that the encoding time for all three context models, which are masked context \cite{DMinnen}, causal context, and causal context plus causal global prediction model. The reason behind this is that the encoding can be performed in parallel while the decoding time for all three models is different as the masked context model takes less time for decoding while the combined causal context and causal global prediction take approximately 39 seconds, as explained in \cite{ZGuo}. The comparison of the encoding and decoding time of all three models is shown in Fig. \ref{EncodingTime}. However, the decoding time of masked context-based CAE architectures used for still image compression is less than the other causal and global context models. The encoding time for all three types of architectures is the same. The use of masked convolution \cite{XHe} has also impacted the visual representation of the reconstructed image. The reconstructed image has better visual quality in \cite{ZGuo} than other state-of-the-art techniques.

Table \ref{Table3} presents the comparison of the performance of the recent deep learning-based learned image compression models.
\begin{table*}
    \centering
     \caption{Summary of the performance of the end-to-end image compression frameworks.}
    \begin{tabular}{p{30pt} p{170pt} p{80pt} p{80pt} p{80pt}}
    \hline
    \hline
         \vspace{0.01cm} \textbf{Research} \vspace{0.1cm}& 
         \vspace{0.01cm} \textbf{Model} \vspace{0.1cm}& 
         \vspace{0.01cm} \textbf{Minimum bpp} \vspace{0.1cm}& 
         \vspace{0.01cm} \textbf{PSNR (dB)} \vspace{0.1cm}& 
         \vspace{0.01cm} \textbf{MS-SSIM} \vspace{0.1cm}\\
         \hline
         \hline
         \vspace{0.01cm} \cite{Theis} \vspace{0.1cm}& 
         \vspace{0.01cm} AE \vspace{0.1cm}&
         \vspace{0.01cm} 0.4 \vspace{0.1cm}& 
         \vspace{0.01cm} 29 \vspace{0.1cm}& 
         \vspace{0.01cm} 0.94 \vspace{0.1cm}\\
         \hline
         \vspace{0.01cm} \cite{Agustsson} \vspace{0.1cm}& 
         \vspace{0.01cm} AE \vspace{0.1cm}&
         \vspace{0.01cm} 0.2 \vspace{0.1cm}& 
         \vspace{0.01cm} -- \vspace{0.1cm}& 
         \vspace{0.01cm} 0.92 \vspace{0.1cm}\\
         \hline
         \vspace{0.01cm} \cite{Zhou} \vspace{0.1cm}& 
         \vspace{0.01cm} VAE \vspace{0.1cm}&
         \vspace{0.01cm} 0.15 \vspace{0.1cm}& 
         \vspace{0.01cm} 30.76 \vspace{0.1cm}& 
         \vspace{0.01cm} 0.955 \vspace{0.1cm}\\
         \hline
         \vspace{0.01cm} \cite{Chen} \vspace{0.1cm}& 
         \vspace{0.01cm} VAE \vspace{0.1cm}&
         \vspace{0.01cm} 0.2 \vspace{0.1cm}& 
         \vspace{0.01cm} 30 \vspace{0.1cm}& 
         \vspace{0.01cm} 0.7768 \vspace{0.1cm}\\
         \hline
         \vspace{0.01cm} \cite{Chen} \vspace{0.1cm}& 
         \vspace{0.01cm} VAE (loss function MSE) \vspace{0.1cm}&
         \vspace{0.01cm} 0.1276 \vspace{0.1cm}& 
         \vspace{0.01cm} 34.63 \vspace{0.1cm}& 
         \vspace{0.01cm} 0.9738 \vspace{0.1cm}\\
         \hline
         \vspace{0.01cm} \cite{Chen} \vspace{0.1cm}& 
         \vspace{0.01cm} VAE (loss function MS-SSIM) \vspace{0.1cm}&
         \vspace{0.01cm} 0.1074 \vspace{0.1cm}& 
         \vspace{0.01cm} 32.54 \vspace{0.1cm}& 
         \vspace{0.01cm} 0.9759 \vspace{0.1cm}\\
         \hline
         \vspace{0.01cm} \cite{Raman} \vspace{0.1cm}& 
         \vspace{0.01cm} CNN \vspace{0.1cm}&
         \vspace{0.01cm} 0.0726 \vspace{0.1cm}& 
         \vspace{0.01cm} 23.93 \vspace{0.1cm}& 
         \vspace{0.01cm} 0.8118 \vspace{0.1cm}\\
         \hline
         \vspace{0.01cm} \cite{Cheng2} \vspace{0.1cm}& 
         \vspace{0.01cm} CNN \vspace{0.1cm}&
         \vspace{0.01cm} 0.519 \vspace{0.1cm}& 
         \vspace{0.01cm} 33.62 \vspace{0.1cm}& 
         \vspace{0.01cm} 0.981 \vspace{0.1cm}\\
         \hline
         \vspace{0.01cm} \cite{Lee} \vspace{0.1cm}& 
         \vspace{0.01cm} CNN \vspace{0.1cm}&
         \vspace{0.01cm} 0.2 \vspace{0.1cm}& 
         \vspace{0.01cm} 31 \vspace{0.1cm}& 
         \vspace{0.01cm} 0.7878 \vspace{0.1cm}\\
         \hline
         \vspace{0.01cm} \cite{Toderici2} \vspace{0.1cm}& 
         \vspace{0.01cm} RNN \vspace{0.1cm}&
         \vspace{0.01cm} 0.5 \vspace{0.1cm}& 
         \vspace{0.01cm} 33.59 \vspace{0.1cm}& 
         \vspace{0.01cm} 0.9877 \vspace{0.1cm}\\
         \hline
         \vspace{0.01cm} \cite{Torfason} \vspace{0.1cm}& 
         \vspace{0.01cm} GAN \vspace{0.1cm}&
         \vspace{0.01cm} 0.0983 \vspace{0.1cm}& 
         \vspace{0.01cm} 28.54 \vspace{0.1cm}& 
         \vspace{0.01cm} 0.973 \vspace{0.1cm}\\
         \hline
         \vspace{0.01cm} \cite{YHu} \vspace{0.1cm}& 
         \vspace{0.01cm} Hyperprior based AE \vspace{0.1cm}&
         \vspace{0.01cm} 0.304 \vspace{0.1cm}& 
         \vspace{0.01cm} 34.4 \vspace{0.1cm}& 
         \vspace{0.01cm} -- \vspace{0.1cm}\\
         \hline
         \vspace{0.01cm} \cite{DMinnenSS} \vspace{0.1cm}& 
         \vspace{0.01cm} Autoregressive Model \vspace{0.1cm}&
         \vspace{0.01cm} 0.01921 \vspace{0.1cm}& 
         \vspace{0.01cm} 18.95 \vspace{0.1cm}& 
         \vspace{0.01cm} -- \vspace{0.1cm}\\
         \hline
         \vspace{0.01cm} \cite{YXie} \vspace{0.1cm}& 
         \vspace{0.01cm} Invertible NN (MSE loss function) \vspace{0.1cm}&
         \vspace{0.01cm} 0.130 \vspace{0.1cm}& 
         \vspace{0.01cm} 31.51 \vspace{0.1cm}& 
         \vspace{0.01cm} 0.972 \vspace{0.1cm}\\
         \hline
         \vspace{0.01cm} \cite{YXie} \vspace{0.1cm}& 
         \vspace{0.01cm} Invertible NN (MS-SSIM loss function) \vspace{0.1cm}&
         \vspace{0.01cm} 0.124 \vspace{0.1cm}& 
         \vspace{0.01cm} 28.01 \vspace{0.1cm}& 
         \vspace{0.01cm} 0.978 \vspace{0.1cm}\\
         \hline
         \vspace{0.01cm} \cite{YChoi} \vspace{0.1cm}& 
         \vspace{0.01cm} Conditional Autoencoder \vspace{0.1cm}&
         \vspace{0.01cm} 0.1697 \vspace{0.1cm}& 
         \vspace{0.01cm} 32.2332 \vspace{0.1cm}& 
         \vspace{0.01cm} 0.9602 \vspace{0.1cm}\\
         \hline
         \vspace{0.01cm} \cite{ZCui} \vspace{0.1cm}& 
         \vspace{0.01cm} Asymmetric Gained VAE (loss function MSE) \vspace{0.1cm}&
         \vspace{0.01cm} 0.107 \vspace{0.1cm}& 
         \vspace{0.01cm} 30.68 \vspace{0.1cm}& 
         \vspace{0.01cm} 0.93 \vspace{0.1cm}\\
         \hline
         \vspace{0.01cm} \cite{ZCui} \vspace{0.1cm}& 
         \vspace{0.01cm} Asymmetric Gained VAE (loss function MS-SSIM) \vspace{0.1cm}&
         \vspace{0.01cm} 0.109 \vspace{0.1cm}& 
         \vspace{0.01cm} 27.76 \vspace{0.1cm}& 
         \vspace{0.01cm} 0.94 \vspace{0.1cm}\\
         \hline
         \hline
    \end{tabular}
    \label{Table3}
\end{table*}

We also compare the run time complexity of the various benchmark learning-driven algorithms with standard codecs. Table \ref{ComparisonTimeComp} shows the comprehensive comparison.
\begin{table}
	\centering
	\caption{Run-time complexity of various learning-driven algorithms with standard codecs.}
	\begin{tabular}{p{60pt}|| p{80pt} p{80pt}}
		\hline
		\hline
		\vspace{0.01cm}\textbf{Codecs}\vspace{0.01cm}&
		\vspace{0.01cm}\textbf{Encoding Time (ms)}\vspace{0.01cm}&
		\vspace{0.01cm}\textbf{Decoding Time (ms)}\vspace{0.01cm}\\
		\hline
		\hline
		\vspace{0.01cm}JPEG\vspace{0.01cm}&
		\vspace{0.01cm}18.600\vspace{0.01cm}&
		\vspace{0.01cm}13.000\vspace{0.01cm}\\
		\hline
		\vspace{0.01cm}JPEG-2000\vspace{0.01cm}&
		\vspace{0.01cm}367.400\vspace{0.01cm}&
		\vspace{0.01cm}80.400\vspace{0.01cm}\\
		\hline
		\vspace{0.01cm}WebP\vspace{0.01cm}&
		\vspace{0.01cm}67.000\vspace{0.01cm}&
		\vspace{0.01cm}83.700\vspace{0.01cm}\\
		\hline
		\vspace{0.01cm}\cite{b15}\vspace{0.01cm}&
		\vspace{0.01cm}242.120\vspace{0.01cm}&
		\vspace{0.01cm}338.090\vspace{0.01cm}\\
		\hline
		\vspace{0.01cm}\cite{JBalle}\vspace{0.01cm}&
		\vspace{0.01cm}64.700\vspace{0.01cm}&
		\vspace{0.01cm}12.100\vspace{0.01cm}\\
		\hline
		\vspace{0.01cm}\cite{ORippel}\vspace{0.01cm}&
		\vspace{0.01cm}8.600\vspace{0.01cm}&
		\vspace{0.01cm}9.900\vspace{0.01cm}\\
		\hline
		\vspace{0.01cm}\cite{DMishra}\vspace{0.01cm}&
		\vspace{0.01cm}3.500\vspace{0.01cm}&
		\vspace{0.01cm}4.000\vspace{0.01cm}\\
		\hline
		\vspace{0.01cm}\cite{CCai}\vspace{0.01cm}&
		\vspace{0.01cm}75.120\vspace{0.01cm}&
		\vspace{0.01cm}73.230\vspace{0.01cm}\\
		\hline
		\vspace{0.01cm}\cite{Cai2018}\vspace{0.01cm}&
		\vspace{0.01cm}79.500\vspace{0.01cm}&
		\vspace{0.01cm}17.400\vspace{0.01cm}\\
		\hline
		\vspace{0.01cm}\cite{Ashok2018}\vspace{0.01cm}&
		\vspace{0.01cm}42.000\vspace{0.01cm}&
		\vspace{0.01cm}32.000\vspace{0.01cm}\\
		\hline
		\vspace{0.01cm}\cite{Li2020}\vspace{0.01cm}&
		\vspace{0.01cm}74.000\vspace{0.01cm}&
		\vspace{0.01cm}984.000\vspace{0.01cm}\\
		\hline
		\vspace{0.01cm}\cite{Li2021}\vspace{0.01cm}&
		\vspace{0.01cm}24.000\vspace{0.01cm}&
		\vspace{0.01cm}32.000\vspace{0.01cm}\\
		\hline
		\vspace{0.01cm}\cite{Toderici2}\vspace{0.01cm}&
		\vspace{0.01cm}1606.900\vspace{0.01cm}&
		\vspace{0.01cm}1079.300\vspace{0.01cm}\\
		\hline
		\hline
	\end{tabular}
	\label{ComparisonTimeComp}
\end{table}
\subsection{Impact of Context-Based Entropy Modeling}
In current learned lossy image compression algorithms, the rate loss is generally the entropy of the codes. To reduce entropy and improve joint rate-distortion performance, a precise estimate of the probabilistic distribution of the codecs is critical. Most of the learning-driven architectures have utilized context-based entropy modeling to improve the RD performance of the codecs. The architectures consider local as well as non-local (NL) contexts. 

The use of global similarity among pixels in image denoising was initially proposed using NL approaches. Deep neural network (DNN)-based image processing approaches then incorporate it into DNNs to use the global information and improve performance in various tasks. The authors in \cite{ABuades} investigated pixel self-similarity and presented the NL means for image denoising based on a content weighted NL average of all pixels in the picture. Wang \textit{et al.} in \cite{XWang} defined the non-local operation as a uniform block, which they used in DNNs to mix local and NL data for object detection. In \cite{DLiu}, Liu \textit{et al.} suggested an NL recurrent network for image restoration that includes NL operations into a recurrent network. The authors of \cite{YZhang} used the NL operation to create attention masks to capture long-range dependence between pixels and pay greater attention to the challenging sections of picture restoration.

Li \textit{et al.} \cite{MLi} proposed learning context to model entropy block in the end-to-end image compression framework. The analysis, as well as synthesis block, utilized U-Net \cite{ORonneberger}-\cite{XLi1} block for encoding and decoding operations. The NL operation was proposed in entropy block to incorporate the global similarities.

All the learning-driven lossy image compression models have limitations and research gaps. The upcoming section highlights the research gaps of learning-driven lossy image compression frameworks.
\section{Research Gaps and Future Directions}\label{Section5}
There are several techniques present for still image compression. All frameworks have their pros and cons. After a detailed survey, we find that there are still research gaps. All ML-based architectures have problems of SRD, parallel acceleration, and OOM.  
\subsection{SRD in Reconstructed Image}
SRD is the striped regions distortion in the reconstructed image using ML-based architectures. SRD in the reconstructed image is one of the open research problems in ML-based image compression techniques. There is no technique available in the literature that addresses this problem. There is a need to solve this problem in future research.
\subsection{Standardization of Architecture}
The second research gap is the standardization of the end-to-end compression architecture. It is very challenging for the researchers. The deep learning architectures present in the literature work well only when the training is performed on GPU and testing is also performed using GPU. Similarly, when the training is done using CPU and testing is performed on CPU, the results are better. However, if the training is performed on the GPU and the testing is done on the CPU and vice versa, the image cannot be reconstructed correctly. There is a requirement for the standard generalized model that gives an optimal solution to this problem.
\subsection{Parallel Acceleration}
The third research gap is due to serial decoding; parallel acceleration of the autoregressive entropy model is impossible. So, there is a requirement for a framework to address the problem of parallel acceleration. In \cite{YWu}, the authors proposed the learned image block-based framework to address this problem. The framework has block artifacts problems for meager bit rates. So, it is still an open research problem. 
\subsection{OOM Problem}
The fourth research gap is given limited GPU resources; full-resolution inference frequently produces OOM problems, especially for high-resolution pictures. Block partition is a good option for dealing with the concerns mentioned above, but it introduces new challenges in terms of decreasing duplication between blocks and removing block effects.
\subsection{Aliasing Effect In Reconstructed Image}
The fifth research gap is the learning-driven lossy image compression frameworks' aliasing effect in the reconstructed image. The reconstructed images of the CNN and CAE-based architectures have variations in the directionality of the patterns, which is called aliasing. This problem needs to be addressed.

These are five major issues in still image compression that can contribute to future research. The upcoming section concludes the survey.
\section{Conclusion}
Images have three types of information, including helpful information, redundant information, and useless information, in the era of the modern world, where the visual quality of images plays a vital role. On the other hand, the is the requirement for more memory to store as well as more bandwidth is required to transmit those high-resolution images. To resolve the problem of memory storage, we perform image compression. Initially, image compression was based on conventional arithmetic and entropy encoding techniques such as JPEG and JPEG-2000. However, since the evolution of ML many several learned image compression models have been proposed in the literature. This paper surveyed several learned image compression techniques based on AEs, VAEs, CNNs, RNNs, GANs, PCA, and fuzzy means clustering. The majority of the architectures are based on CNNs and AEs. These models achieve better compression efficiency than JPEG-2000, which is an anchor in image compression. We also highlighted four significant research gaps in ML-based image compression models. Those gaps are the SRD problems, architecture standardization, parallel acceleration, aliasing, and OOM. These problems are yet to be addressed in still image compression.



\end{document}